\documentclass{aa}  

\usepackage{graphicx}
\usepackage{xcolor}

\usepackage{txfonts}
\begin{document} 

    \title{Cosmological constraints on the magnification bias on sub-millimetre galaxies after large-scale bias corrections.}
    \titlerunning{Cosmology with magnification bias of SMGs}
    \authorrunning{Gonzalez-Nuevo J. et al.}

    \author{Gonz{\'a}lez-Nuevo J.\inst{1,2}, Cueli M. M.\inst{1,2}, Bonavera L.\inst{1,2}, Lapi A.\inst{3,4}, Migliaccio M.\inst{5,6},\\Arg{\"u}eso F. \inst{7,2}, Toffolatti L.\inst{1,2}}

   \institute{$^1$Departamento de F{\'i}sica, Universidad de Oviedo, C. Federico Garcia Lorca 18, 33007 Oviedo, Spain\\
              $^2$Instituto Universitario de Ciencias y Tecnolog{\'i}as Espaciales de Asturias (ICTEA), C. Independencia 13, 33004 Oviedo, Spain\\
              $^3$International School for Advanced Studies (SISSA), via Bonomea 265, I-34136 Trieste, Italy\\
              $^4$Institute for Fundamental Physics of the Universe (IFPU), Via Beirut 2, I-34014 Trieste, Italy\\
              $^5$Dipartimento di Fisica, Universit\`a di Roma Tor Vergata, Via della Ricerca Scientifica, 1, Roma, Italy\\
              $^6$INFN, Sezione di Roma 2, Universit\`a di Roma Tor Vergata, Via della Ricerca Scientifica, 1, Roma, Italy\\
              $^7$Departamento de Matem{\'a}ticas, Universidad de Oviedo, C. Federico Garcia Lorca 18, 33007 Oviedo, Spain\\
}

   \date{Received xxx, xxxx; accepted xxx, xxxx}

 
  \abstract
   {The study of the magnification bias produced on high-redshift sub-millimetre galaxies by foreground galaxies through the analysis of the cross-correlation function was recently demonstrated as an interesting independent alternative to the weak-lensing shear as a cosmological probe.} 
   {In the case of the proposed observable, most of the cosmological constraints mainly depend on the largest angular separation measurements. Therefore, we aim to study and correct the main large-scale biases that affect foreground and background galaxy samples to produce a robust estimation of the cross-correlation function. Then we analyse the corrected signal to derive updated cosmological constraints.}
   {We measured the large-scale, bias-corrected cross-correlation functions using a background sample of H-ATLAS galaxies with photometric redshifts > 1.2 and two different foreground samples (GAMA galaxies with spectroscopic redshifts or SDSS galaxies with photometric ones, both in the range 0.2 < z < 0.8). These measurements are modelled using the traditional halo model description that depends on both halo occupation distribution and cosmological parameters. We then estimated these parameters by performing a Markov chain Monte Carlo under multiple scenarios to study the performance of this observable and how to improve its results.}
   {After the large-scale bias corrections, we obtain only minor improvements with respect to the previous magnification bias results, mainly confirming their conclusions: a lower bound on $\Omega_m > 0.22$ at $95\%$ C.L. and an upper bound $\sigma_8 < 0.97$ at $95\%$ C.L. (results from the $z_{spec}$ sample). Neither the much higher surface density of the foreground photometric sample nor the assumption of Gaussian priors for the remaining unconstrained parameters significantly improve the derived constraints. However, by combining both foreground samples into a simplified tomographic analysis, we were able to obtain interesting constraints on the $\Omega_m$-$\sigma_8$ plane as follows: $\Omega_m= 0.50_{- 0.20}^{+ 0.14}$ and $\sigma_8= 0.75_{- 0.10}^{+ 0.07}$ at 68\% CL.}
   {}

   \maketitle
%

\section{Introduction}

The apparent excess number of high-redshift sources observed near low-redshift mass structures is known as magnification bias \citep[see e.g.][]{SCH92}. The deflections produced by the intervening gravitational field (area stretching and amplification) affecting the light rays coming from distant sources, in general, increase their chances of being included in a flux-limited sample \citep[see e.g.][]{ARE11}.

An unambiguous manifestation of this bias is the existence of a non-negligible cross-correlation function between two source samples with non-overlapping redshift distributions. The magnification bias has been observed in several contexts: a galaxy-quasar cross-correlation function \citep{SCR05, MEN10}, a cross-correlation signal between Herschel sources and Lyman-break galaxies \citep{HIL13}, or the cosmic microwave background \citep[CMB; ][]{BIA15, BIA16} among others.

The cross-correlation signal can be enhanced by optimising the choice of foreground and background samples. In this paper we use  sub-millimetre galaxies (SMGs) as the background sample because some of their features (steep luminosity function, very faint emission in the optical band, and typical redshifts above $ z > 1-1.5 $) make them close to the optimal background sample for lensing studies as confirmed by a long series of publications \citep[see for example][among the most important ones]{BLA96, NEG07, NEG10, GON12,BUS12,BUS13,FU12, WAR13, CAL14, NAY16, NEG17, GON19, BAK20}.

In early works, the magnification bias produced on SMGs was already observed \citep{WAN11} and measured with high significance of  $> 10\sigma$ \citep{GON14}. In \citet{GON17} the measurements were improved, facilitating a more detailed study with a halo model. It was concluded that the lenses are massive galaxies or even galaxy groups or clusters that have a minimum mass of $M_{lens}\sim10^{13}M_{\odot}$. Moreover, it was demonstrated that it is possible to split the foreground sample into different redshift bins and to perform a tomographic analysis thanks to better statistics.
Finally, \citet{BON19} used the magnification bias to study the mass properties of a different type of lenses, a sample of quasi-stellar objects (QSOs) at $0.2<z<1.0$. It was possible to estimate the halo mass where the QSOs acting as lenses are located in the sky, $M_{min} = 10^{13.6_{-0.4}^{+0.9}} M_{\odot}$. These mass values indicate that we are observing the lensing effect of a cluster size halo signposted by the QSOs.

The interest in magnification bias is driven by the fact that it can be used as an additional cosmological probe to address the estimation of the parameters in the standard cosmological model. The importance of the magnification bias effect depends on the gravitational deflection caused by low-redshift galaxies on light travelling close to such lens, which in turn depends on cosmological distances and galaxy halo properties.

Features such as the anisotropies in the CMB \citep[e.g.][]{HIN13, PLA16_XIII, PLA18_VI}, the big bang nucleosynthesis \citep[e.g.][]{FIE06}, and the SNIa observations of the Universe accelerating expansion \cite[e.g.][]{BET14} are handled well by the current standard cosmological model. This model also includes some large-scale structure (LSS) significant predictions about galaxies distributions (e.g. \cite{PEA01}) such as baryon acoustic oscillations (BAOs) (e.g. \cite{ROS15}). Therefore, measurements based on such observables provides independent and complementary constraints on the cosmological parameters \citep[e.g.][]{PEA94}. The current model is successful because results from different probes are in great agreement.

However, with the increase in the quality and quantity of the measurements, some tensions and small-scale issues have arisen that might indicate the necessity of modifications of the $\Lambda$CDM model. The main tensions are the value of the Hubble constant, $H_0$ \citep[$74.03 \pm 1.42$ km/s/Mpc by][ with $67.4 \pm 0.5$ km/s/Mpc]{RIE19, PLA18_VI} and the usually degenerate relationship between the $\Omega_m$ and $\sigma_8$ parameters \citep[e.g.][]{HEY13,PLA16_XXIV,HIL17, PLA18_VI}.

In this context, \citet{BON20} (hereafter BON20) test the capability of the magnification bias produced on high-z SMGs as an additional independent cosmological probe in the effort to resolve the tensions. With this proof of concept analysis, $\Omega_m$ and $H_0$ are not well constrained. However, interesting limits are found: a lower limit of $\Omega_m>0.24$ at 95\% CL and an upper limit of $\sigma_8<1.0$ at 95\% CL (with a tentative peak around 0.75).

Although the derived cosmological constraints from the magnification bias are relatively weak, this bias was confirmed as a new, independent observable, thereby making it a valuable new technique. Therefore, it is worth making an effort to improve the results.

In this respect, most of the cosmological analysis that can be performed using the measured cross-correlation function (e.g. cosmological parameters, mass function, and neutrinos) depends mainly on the observed data at the largest angular scales ($\gtrsim20$ arcmin). On the one hand, these data are the most uncertain with large error bars. Large areas and high source densities are needed to derive precise measurements. On the other hand, large-scale bias, which can be considered negligible at smaller scales, can affect the data, and, as a consequence, the derived cosmological results. For these reasons the main goal of this work is to deeply study and find the optimal strategy to measure and analyse a precise and unbiased cross-correlation function at cosmological scales.

The work is organised as follows. In section \ref{sec:data} the background and foreground samples are described and in section \ref{sec:methodology} the methodology is presented. The large-scale biases and how to correct them are described in \ref{sec:LS_bias}. The derived cosmological constraints and conclusions are discussed in sections \ref{sec:results} and \ref{sec:conclusion}, respectively. In Appendix \ref{sec:corner_plots} we show the posteriors distributions of all the cases analysed and discussed in this work.

\section{Data}
\label{sec:data}
The different galaxy samples used in this work are described in this section: the background sample, consisting of SMGs sources; and the foreground samples, consisting of two independent samples with spectroscopic and photometric redshifts, respectively.

\begin{figure}[ht]
\centering
\includegraphics[width=\columnwidth]{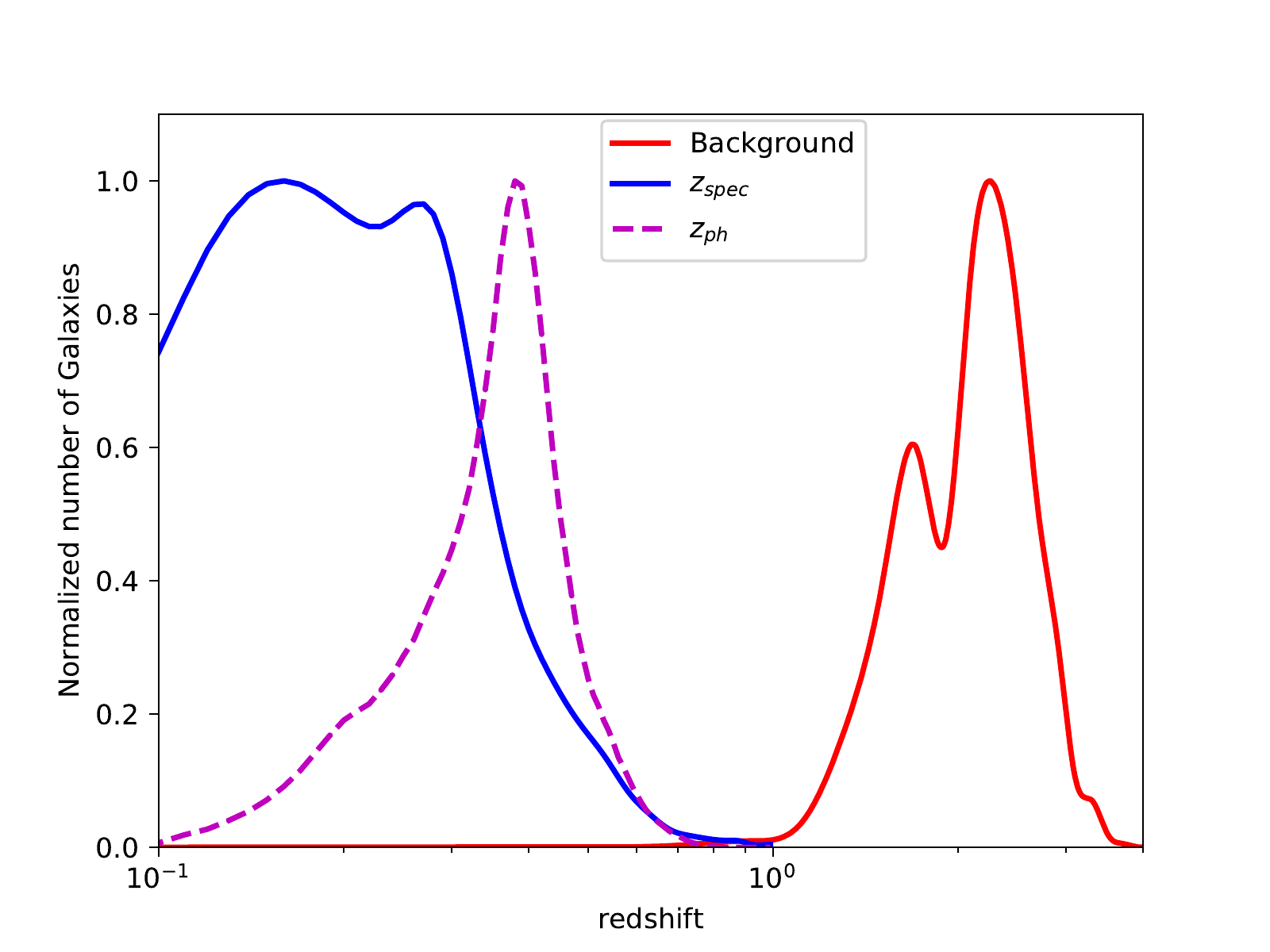}
    \caption{Normalised redshift distributions of the three catalogues used in this work: the background sample, that is H-ATLAS high-z SMGs (solid red line); the GAMA spectroscopic foreground sample (solid blue line); and the SDSS photometric foreground sample (dashed magenta line).}
    \label{Fig:dNdz_hist}
\end{figure}

\subsection{Background sample}
The Herschel Astrophysical Terahertz Large Area Survey \citep[H-ATLAS;][]{EAL10} is the largest area extragalactic survey carried out by the Herschel space observatory \citep{PIL10} covering  $\sim 610 deg^2$ with the Photodetector Array Camera and Spectrometer \citep[PACS;][]{POG10} and the Spectral and Photometric Imaging REceiver \citep[SPIRE;][]{GRI10} instruments between 100 $\mu m$ and 500 $\mu m$. Details of the H-ATLAS map-making, source extraction, and catalogue generation can be found in \citet{IBA10, PAS11, RIG11, VAL16, BOU16}, and \citet{MAD20}.

The background sample consists of H-ATLAS sources detected in the three Galaxy And Mass Assembly (GAMA) survey fields (total common area of $\sim 147 deg^2$), the north Galactic pole (NGP; $\sim 170 deg^2$) and the part of the south Galactic pole (SGP) that overlaps with the spectroscopic foreground sample ($\sim 60 deg^2$). A photometric redshift selection of 1.2 < z < 4.0 was applied to ensure no overlap in the redshift distributions of lenses and background sources, and we are thus left with $\sim66000$ ($\sim 24$ per cent of the initial sample and $z_{ph,med} = 2.20$). The redshifts estimation is described in detail in \citet{GON17,BON19} and references therein. This is the same background sample used in \citet{GON17}, \citet{BON19}, and BON20.

\subsection{Foreground samples}
In this work we use two independent foreground samples. The first is  BON20, which is the same sample used by \citet{GON17}; we call this sample the "$z_{spec}$ sample". It consists of a sample extracted from the GAMA II \citep{DRI11, BAL10,BAL14,LIS15} spectroscopic survey and has $\sim 150000$ galaxies in the common area for 0.2 < $z_{spec}$ < 0.8 ($z_{spec,med} = 0.28$).

The H-ATLAS and GAMA II surveys were carried out to maximise the common area coverage. Both surveys covered the three equatorial regions at 9, 12, and 14.5 h (referred to as G09, G12, and G15, respectively), but, as a consequence
of the scanning strategy, the overlap is not perfect (as just a few percent of these regions are missing). Only the common areas were taken into account (total common area of $\sim 147 deg^2$). The SGP region was only partially observed by GAMA II as well ($\sim 60 deg^2$ as stated before). The NGP region was not covered by the GAMA II survey and, therefore, was not used for the $z_{spec}$ sample.  Thus, the resulting total common area is of about $\sim 207 deg^2$, surveyed down to a limit of $r \simeq 19.8$ mag.
This is the same foreground sample used in \citet{GON17} and BON20.

The second foreground sample was selected from the 16th data release of the Sloan Digital Sky Survey \citep[SDSS; ][]{BLA17,AHU19}. This sample consists of galaxies with photometric redshift between 0.2< $z_{ph}$ < 0.8 and photometric redshift error $z_{err}/(1+z)<1$ (\textit{photoErrorClass}=1). The SDSS has completely covered the H-ATLAS equatorial regions and the NGP region (a total common area of $\sim 317 deg^2$). The SGP region was not covered by SDSS and, therefore, was not used for this sample. Thus, the second foreground sample, denominated "$z_{ph}$ sample", comprises $\sim962000$ galaxies in total in the common area with median value of $z_{ph,med} = 0.38$.

The reason to introduce this second foreground sample is to study the improvements in the final results by increasing the density of potential lenses. The higher uncertainty in the redshift estimation of the foreground photometric redshifts is not very important in the current analysis because we are using a single wide redshift bin.

The normalised redshift distributions of the different samples are compared in Figure \ref{Fig:dNdz_hist}. As in \citet{GON17}, the random errors in the photometric redshifts are taken into account to estimate the redshift distributions. The main effect is to broaden the distributions beyond the selection limits.
Figure \ref{Fig:dNdz_hist} clearly shows the gap in redshift between the background and the foreground sources. The same figure also highlights the different redshift distributions between the two foreground samples.

\section{Methodology}
\label{sec:methodology}

\subsection{Tiling area scheme}
\begin{figure}[ht]
\centering
\includegraphics[width=\columnwidth]{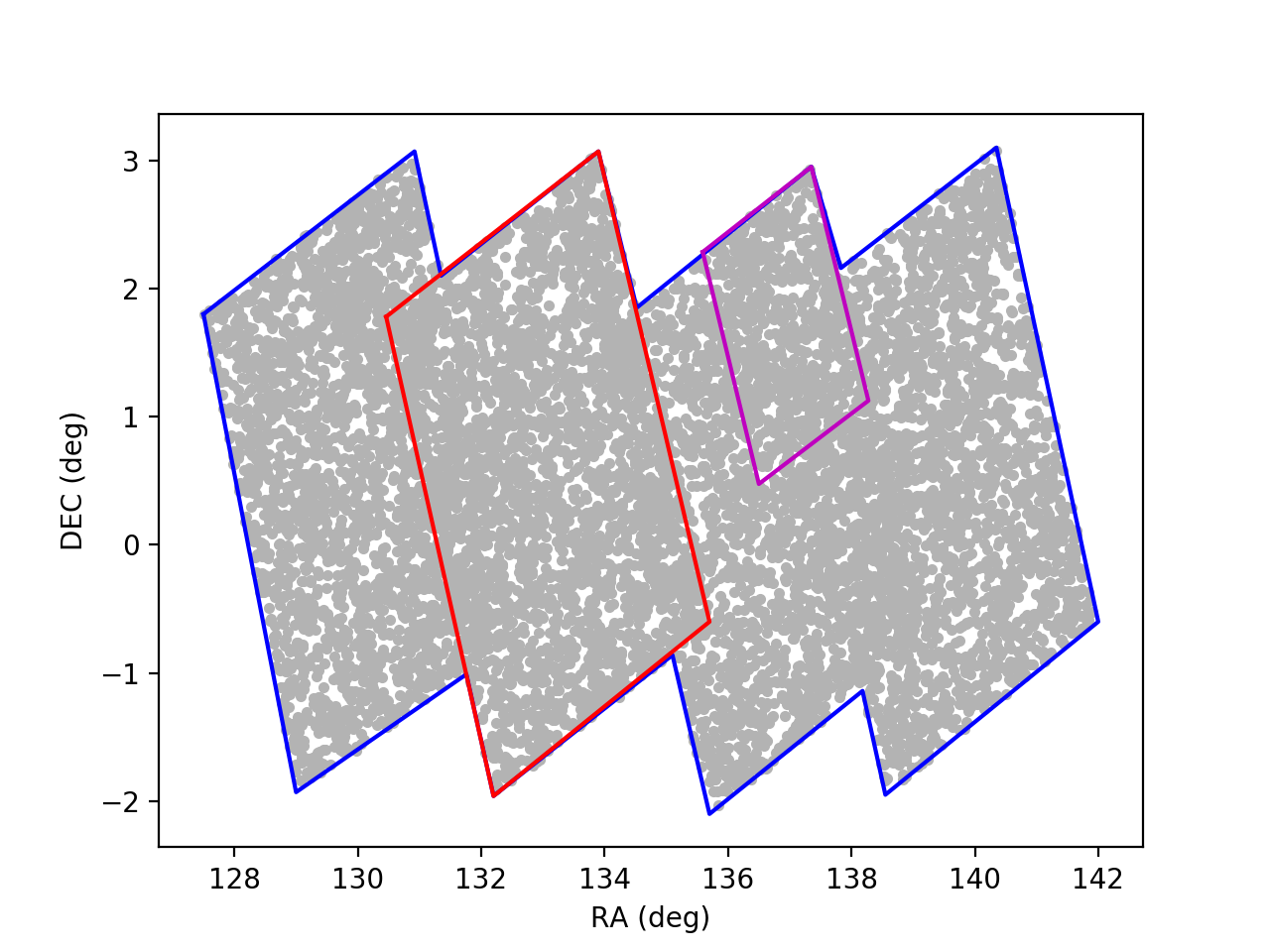}
    \caption{Examples of the different area selection to measure the cross-correlation function for the G09 H-ATLAS field. The All field area is shown in blue (~56 sq. deg). The Tile selection is shown in red (~4x4 sq. deg.) and the mini-Tile area in magenta (~2x2 sq. deg.)
 }
    \label{Fig:Tiles}
\end{figure}

The H-ATLAS survey is divided in five different fields: three GAMA fields in the ecliptic (9h, 12h, 15h) and two in the NGP and SGP. The H-ATLAS scanning strategy produced the characteristic diamond repeated shape in most of their fields \ref{Fig:Tiles}. Taking into account the available area in each field we have different possibilities to measure the cross-correlation function.

The "All" field area (blue line) provides the best statistics (i.e. smaller statistical uncertainties) both at small and large scales. The drawback is that we are limited to four to five fields to minimise the cosmic variance.

The "Tile" area (red line) is the straightforward shape to be selected, taking into account the observational strategy. The area of each tile\ (16 sq. deg) should be large enough to avoid a bias in the large-scale measurements (normally limited to angular separation below 2 deg). To maintain a regular shape for the tiles, a small overlap among such regions is needed, which is typically lower than 20\% of the area of the tiles. The advantage of this area scheme is in the fact that it provides around 24 different tiles, which should help to diminish the cosmic variance.

The "mini-Tile" area (magenta line) is built by dividing the tiles into four equal mini-Tile areas (each of ~2x2 sq. deg). This area scheme typically provides around 96 different tiles. However, the maximum distance allowed by this area scheme is close to the cosmological scales that we want to measure. This was the area scheme used in BON20.

Each tiling area scheme has its own strong and weak points and can be affected by different types of large-scale biases. Therefore, we performed a detailed analysis to compare the measurements from the different tiling area schemes and derived a robust estimation of the cross-correlation function, in particular at the cosmological angular scales.

\subsection{Angular cross-correlation function estimation}
\label{sec:cross_corr}

As described in detail in \cite{GON17}, BON20, we used a modified version of the \cite{LAN93} estimator \citep{HER01} as follows:
\begin{equation}
\label{eq:wx}
w_x(\theta)=\frac{\rm{D}_1\rm{D}_2-\rm{D}_1\rm{R}_2-\rm{D}_2\rm{R}_1+\rm{R}_1\rm{R}_2}{\rm{R}_1\rm{R}_2}
,\end{equation}
where $\rm{D}_1\rm{D}_2$, $\rm{D}_1\rm{R}_2$, $\rm{D}_2\rm{R}_1$ and $\rm{R}_1\rm{R}_2$ are the normalised data1-data2, data1-random2, data2-random1 and random1-random2 pair counts for a given separation $\theta$.

For each selected area, we computed the angular cross-correlation function and the statistical error. For stability, we averaged between ten different realisations using different random catalogues each time. The uncertainty related to the random realisation depends on the area size, being worse for smaller areas. For the smallest areas used in this work, there is typically a $\sim12$\% variation on the measured mean. However, even in this case, this variation only corresponds to a $\sim11$\% of the cosmic variance, that is the mean variation between different sky areas.
To minimise the cosmic variance, each final measurement corresponds to the mean value of the cross-correlation functions estimated in each individual selected area for a given angular separation bin. The uncertainties correspond to the standard error of the mean, that is $\sigma_\mu=\sigma/\sqrt{n}$ with $\sigma$ the standard deviation of the population and $n$ the number of independent areas; each selected region can be assumed as statistically independent owing to the small overlap.
We tested that using alternative statistical approaches to estimate the uncertainties, for example jackknife, provides equivalent results.

\begin{figure*}[ht]
\centering
\includegraphics[width=\columnwidth]{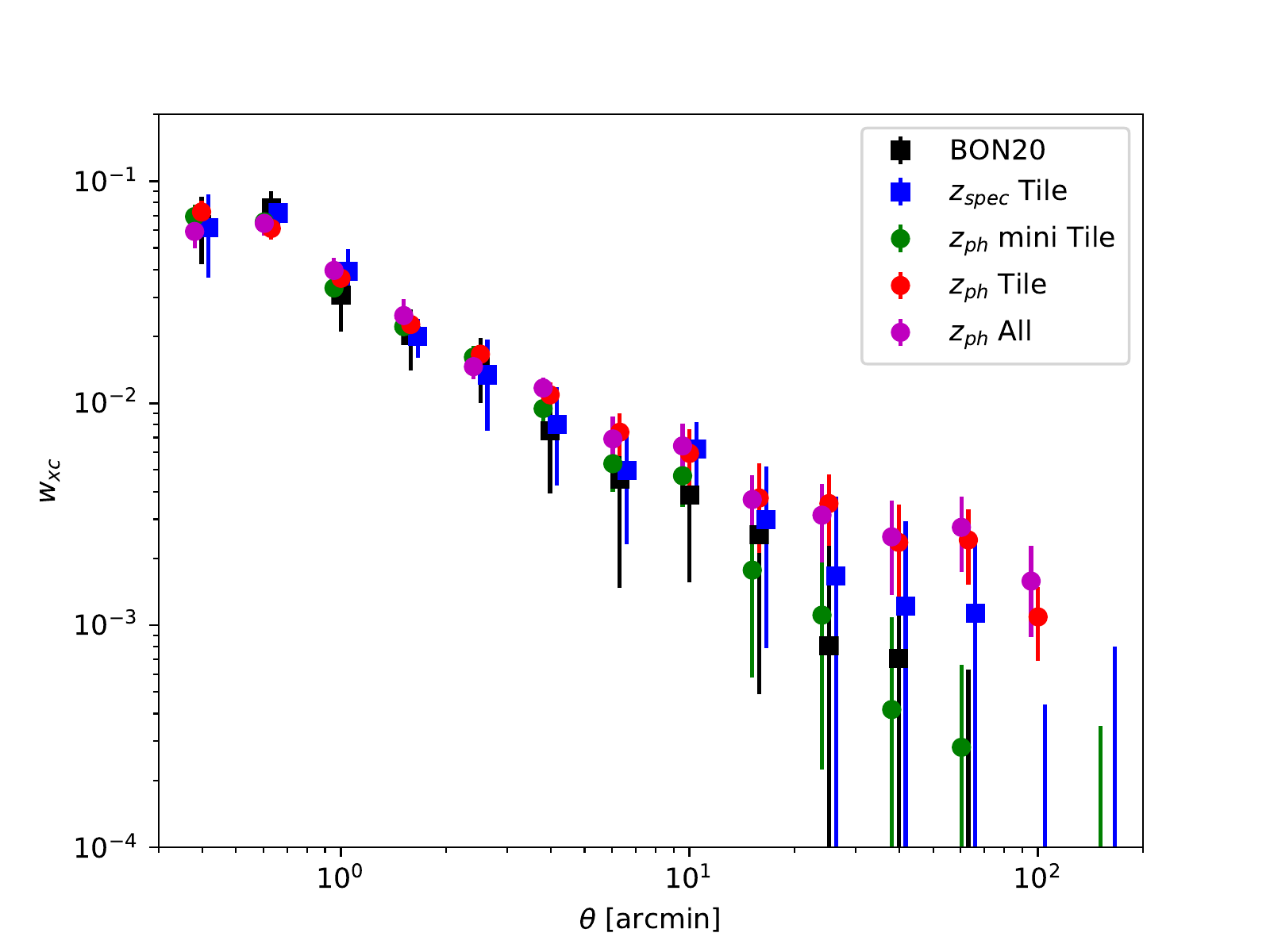}
\includegraphics[width=\columnwidth]{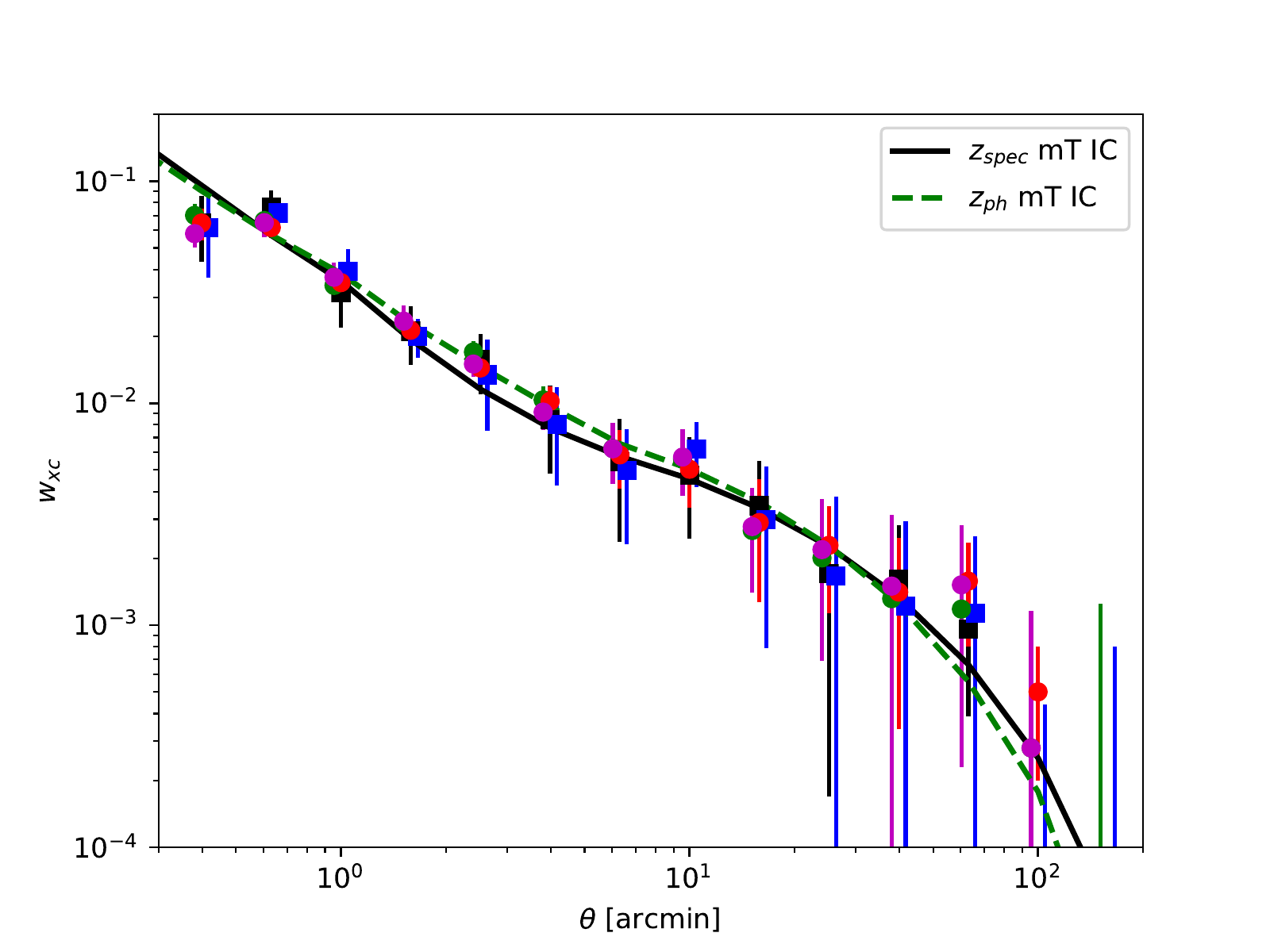}
    \caption{Cross-correlation measurements for the $z_{spec}$ and $z_{ph}$ foreground samples using the mini-Tile and Tile schemes. Left panel: Measurements before any large-scale bias correction. The $z_{spec}$ mini-Tile case exactly matches the measurements analysed in BON20.  Right panel: The same measurements after the corrections are applied. The model predictions using the best-fit values for the $z_{spec}$ sample (solid black line) and the $z_{ph}$ sample (dashed green line) in the mini-Tile scheme are also shown (see text for more details).}
    \label{Fig:xcorr_data}
\end{figure*}

\subsection{Halo model}
As described in detail in the above-mentioned works \citep{GON17,BON19}, in BON20 we adopted the halo model formalism proposed by \citet{COO02} to interpret a foreground-background source cross-correlation signal. A halo is defined as spherical regions whose mean over-density with respect to the background at any redshift is given by its virial value, which is estimated following \cite{WEI03} assuming a flat $\Lambda$CDM model. We used the traditional \cite{NAV96} density profile with the concentration parameter given in \cite{BUL01}.

The cross-correlation between the foreground and background sources is linked to the low-redshift galaxy-mass correlation through the weak gravitational lensing effect.
The foreground galaxy sample traces the mass density field that causes the weak lensing, affecting the number counts of the background galaxy sample through magnification bias.

Following mainly \cite{COO02} \citep[see][for details]{GON17}, we compute the correlation between the foreground and background sources adopting the standard Limber \citep{LIM53} and flat-sky approximations \citep[see e.g.][and references therein]{KIL17}. It can be estimated as
\begin{equation}
    \begin{split}
    w_{fb}=2(\beta -1)\int^{z_s}_0 \frac{dz}{\chi^2(z)}\frac{dN_f}{dz}W^{lens}(z) \\
    \int_{0}^{\infty}\frac{ldl}{2\pi}P_{gal-dm}(l/\chi^2(z),z)J_0(l\theta) 
    \end{split}
,\end{equation}
where
\begin{equation}
    W^{lens}(z)=\frac{3}{2}\frac{H_0^2}{c^2}E^2(z)\int_z^{z_s} dz' \frac{\chi(z)\chi(z'-z)}{\chi(z')}\frac{dN_b}{dz'}
,\end{equation}
where $E(z)=\sqrt{\Omega_m(1+z)^3+\Omega_{\Lambda}}$, $dN_b/dz$ and $dN_f/dz$ as the unit-normalised background and foreground redshift distribution and $z_s$ the source redshift. $\chi(z)$ is the comoving distance to redshift z. The logarithmic slope of the background sources number counts is assumed $\beta=3$ ($N(S) = N_0 S^{-\beta}$) as in previous works \citep{LAP11, LAP12, CAI13, BIA15, BIA16, GON17, BON19}. Small variations of its value are almost completely compensated by small changes in the $M_{min}$ parameter.

As the halo occupation distribution (HOD), we adopted the three parameters from the \citet{ZHE05} model. In this model, all haloes above a minimum mass $M_{min}$ host a galaxy at their centre, while any remaining galaxy is classified as satellite. Satellites are distributed proportionally to the halo mass profile and haloes host them when their mass exceeds the $M_1$ mass. Finally, the number of satellites is a power-law function of halo mass with $\alpha$ as the exponent, $N_{sat}(M)=(\frac{M}{M_1})
^\alpha$. Therefore, $M_\text{min}$, $M_1$ and $\alpha$ are the astrophysical free-parameters of the model.

\subsection{Estimation of parameters}
\label{sec:par_estimate}

To estimate the different set of parameters, we performed a Markov chain Monte Carlo (MCMC) using the open source {\it emcee} software package \citep{EMCEE}. It is a pure-Python implementation of \cite{GOO10} affine invariant MCMC ensemble sampler licensed by the Massachusetts Institute of Technology. For each run, we generated at least 90000 posterior samples to ensure a good statistical sampling after convergence.

In the cross-correlation function analysis, we took into account both the astrophysical HOD parameters, and the cosmological parameters. The astrophysical parameters to be estimated are $M_{min}$, $M_1$, and $\alpha$. The cosmological parameters we want to constrain are $\Omega_m$, $\sigma_8$ and $h=H_0/100$. With the current samples, we do not have the statistical power to constrain $\Omega_B$, $\Omega_{\Lambda}$, and $n_s$ in our analysis. As we assume a flat universe, $\Omega_{\Lambda}$ is simply: $\Omega_{\Lambda}=1-\Omega_m$. For the the other two cosmological parameters, we keep them fixed to the \textit{Planck} most recent results $\Omega_B=0.0486$ and $n_s=0.9667$ (see \citet{PLA18_VI}).

A traditional Gaussian likelihood function was used in this work.
It should be noted that only the cross-correlation data in the weak-lensing regime ($\theta \ge 0.2$ arcmin) are taken into account for the fit since we are in the weak-lensing approximation \citep[see][for a detailed discussion]{BON19}. 

In general, we used the same flat priors for all the different analyses. These are based on those used in BON20. As for the astrophysical parameters, we chose [12.0-13.5] for $\log M_{min}$, [13.0-15.5] for $\log M_1$ and [0.5-1.5] for $\alpha$. For the cosmological parameters, we chose [0.1-0.8] for $\Omega_m$, [0.6-1.2] for $\sigma_8$ and [0.5-1.0] for $h$.

\section{Large-scale biases}
\label{sec:LS_bias}
The cross-correlation function measurements using the different tiling area schemes are compared in Figure \ref{Fig:xcorr_data}. The left panel shows the measurements before any correction is applied. While all the measurements agree almost perfectly within the uncertainties at small scales, there is a widespread variation of estimated values for angular separations above $\sim 10$ arcmin. But the cosmological parameters affect mainly those angular scales (see BON20 appendix figures). Therefore, we need to understand the causes that produce such high variation on our observations at those large angular scales before attempting any robust cosmological analysis.

It is well known that the distribution of galaxies in the Universe is not perfectly homogeneous. Therefore, in a field with a limited area, the number of detected galaxies is somewhat higher or lower than the mean value obtained considering large enough areas. If this variation is not taken into account when building the random catalogues for a particular field, it affects data-random (DR) and random-random (RR) related terms in equation \ref{eq:wx} and the estimated correlation could be stronger or weaker than the intrinsic value (see e.g. \citet{ADE05} for a detailed discussion on this topic).
To this respect, there are mainly two different biases that can affect the cross-correlation measurements at large scales: the integral constraint \citep[IC;][]{ROC99} and the surface density variation \citep[][]{BLA02}. 

\subsection{Integral constraint }
 When many fields are averaged, the overall effect of the large-scale fluctuations tends to make the observed correlation weaker mainly at the largest observed scales. This means that the estimated cross-correlation function is biased low by a constant, the IC $w_{x\_ideal}(\theta) = w_{x}(\theta) + IC$.
 
 Although there are possible theoretical approaches to estimate the IC for a particular scanning strategy (see e.g. \citet{ADE05}), it is commonly estimated numerically using the following RR counts:
 \begin{equation}
     IC=\frac{\sum_i{\rm{R}_1\rm{R}_2(\theta_i)w_{x\_ideal}(\theta_i)}}{\sum_i{\rm{R}_1\rm{R}_2(\theta_i)}}.
 \end{equation}

As a first approximation of $w_{x\_ideal}(\theta_i)$, we assumed a power-law model, $w_{x\_ideal}(\theta_i)= A \theta^\gamma$. In order to be as independent as possible of the exact value of the cosmological parameters (that mainly affect the largest angular scales), we estimated the best-fit parameters for the power law using only the observed cross-correlation function below 20 arcmin ($A=10
^{-1.54}$ arcmin and $\gamma=-0.89$). With the estimated power law, the derived IC value for the mini-Tiles area was $9\times10^{-4}$.
We verified that choosing a smaller angular separation upper limit or using different data sets did not affect the derived IC value.

Moreover, assuming the best-fit model of BON20, which can be considered biased low because they neglected the IC correction, the derived IC is again of the same value. Therefore, we can conclude that the mini-Tiles estimated cross-correlation functions at the largest scales (>20 arcmin) are biased low, but can be safely corrected by adding an $IC = 9\times10^{-4}$. Anyway, as discussed in section \ref{sec:flat_priors}, this correction does not introduce any substantial difference with respect to the BON20 results on cosmological parameters.

On the other hand, the estimated IC for the Tiles area is $IC = 5\times10^{-4}$, considering both the power-law fit and the BON20 best-fit model. As expected, the correction is smaller than in the mini-Tiles case, taking into account the larger area of the Tiles. The IC in the Tiles case only marginally affects the measurements above $\sim 40$ arcmin. Considering the large uncertainties at those angular scales, it can be almost considered a negligible correction for the $z_{spec}$ sample measured using the Tiles area. However, to be precise, we decided to apply it in any case.

On the other hand, the IC results are completely negligible in the case of using the All field area scheme, as expected.

\subsection{Surface density variations}
The results using the Tiles area differ for the $z_{spec}$ and $z_{ph}$ samples. This difference remains after the IC correction because it is the same for both cases. Moreover, the discrepancy is even stronger in the All scenario case (since the All measurements are almost the same between both samples, we are focussing only on the $z_{ph}$ for simplicity). This is a clear indication that an additional large-scale bias is affecting the measurements when larger areas are considered. The fact that the $z_{ph}$ sample is more affected is probably related to the much higher density of sources in this sample.

If there is an additional variation of the source density of the foreground or the background sample that is not taken into account when building the random catalogues, it can produce a spurious enhancement of the measured correlation. As explained by \citet{BLA02}, the number of close pairs depended on the local surface density while the random pairs are related to the global average surface density. Then, systematic fluctuations produce $DD>RR$, which means a higher correlation (e.g. consider just the simplest estimator of the auto-correlation: $w(\theta)=DD/RR-1$). Therefore, if present, the surface density variation produces the opposite effect with respect to the IC, that is what we are observing with the $z_{ph}$ sample.

\subsubsection{Instrumental noise variation}
\begin{figure}[ht]
\centering
\includegraphics[width=\columnwidth]{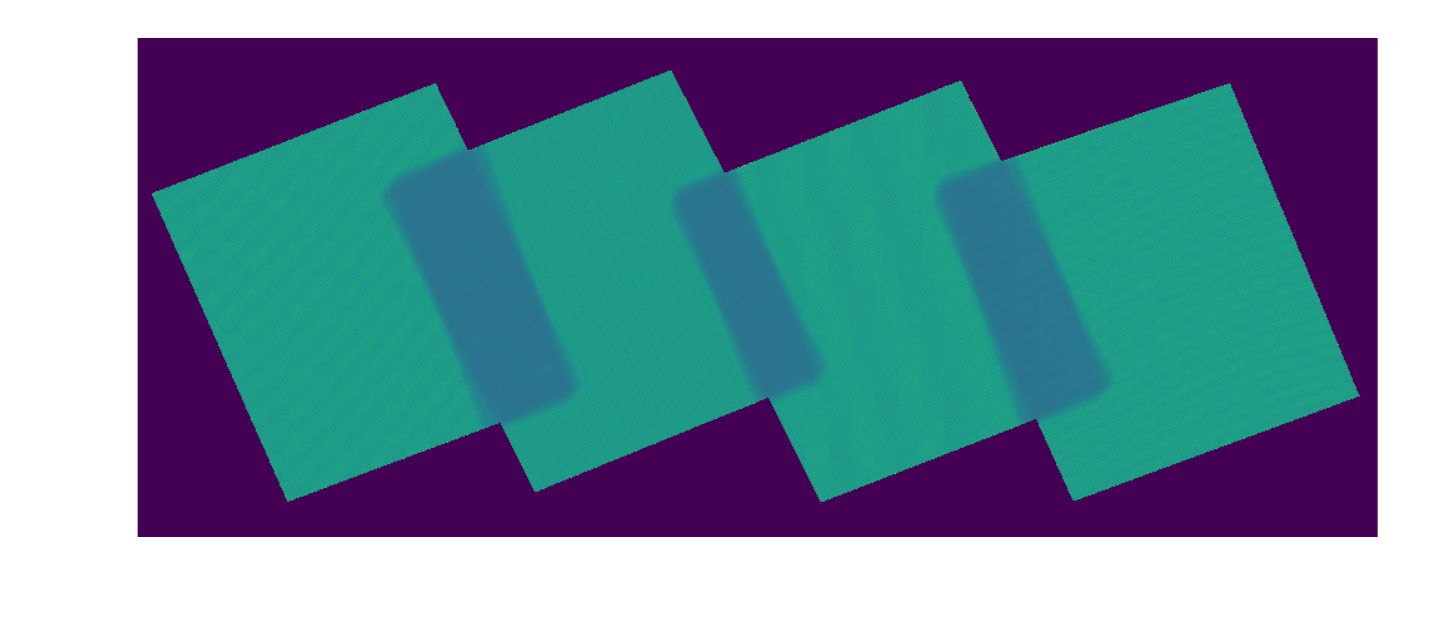} \\
\includegraphics[width=\columnwidth]{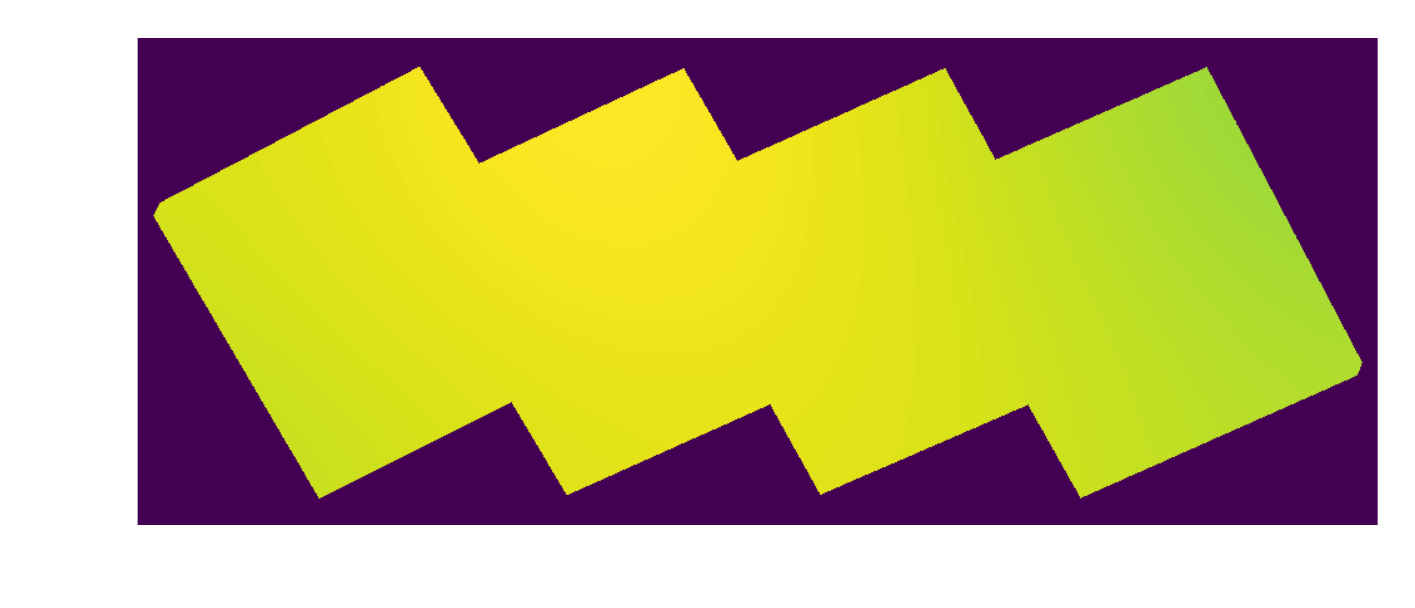}
    \caption{Top panel: Example of the instrumental noise variation in the G09 field due to the scanning strategy. Bottom panel: Example of the surface density variation for the $z_{ph}$ sample in the G15 filed after been filtered using a Gaussian kernel with a standard deviation of 180 arcmin.}
    \label{Fig:SD_maps}
\end{figure}

For the background sample, there is a well-known surface density variation related to the instrumental noise due to the scanning strategy (see Figure \ref{Fig:SD_maps}, top panel). The overlap between the tiles reduces the instrumental noise, which allows fainter SMGs to be detected with respect the rest of the field. For the auto-correlation analysis it was demonstrated that the potential effect can be considered negligible \citep{AMV19}. Moreover, our results indicate that the relatively low surface density of the $z_{spec}$ sample also makes this effect negligible. In other words, the number of additional pairs due to the fainter background sources in those areas is not relevant enough to affect the measurements for the $z_{spec}$ sample. However, the much higher surface density of the $z_{ph}$ sample could produce a relevant enough enhancement of background-foreground pairs in those regions, therefore, inducing a large-scale surface density variation for the Tiles and All area schemes; we can consider the mini-Tile measurements simply dominated by the IC correction and neglect this other type of large-scale bias even for the $z_{ph}$ sample.

To correct the instrumental noise surface density bias, we adopted the same procedure to generate random catalogues used in \citet{AMV19} for the auto-correlation analysis of the SMGs. First, a flux was randomly chosen among the flux densities of our background sample. Then the simulated galaxy is situated in a random position of the field. At this position the local noise was estimated as the instrumental noise and the confusion noise \citep[see Table 3 of][for the GAMA fields]{VAL16}. The estimated local noise is used to introduce a random Gaussian perturbation in the flux density. Finally, the simulated galaxy was kept in the sample if its flux density was greater than four times the local noise, the same detection limit used to produce the official H-ATLAS catalogue. This process was repeated for each random galaxy until the completion of the random catalogue.
These newly generated random catalogues only correspond to the background sample, that is it was only applied to build the $R_1$ random catalogues (used to estimate the $D_2R_1$ and $R_1R_2$ terms).

When the instrumental noise variation is considered, the cross-correlation functions showed a small correction towards lower values at the largest angular scales (not shown individually in Figure \ref{Fig:xcorr_data}). Although this result confirms that this bias is not negligible, it also highlights that it is not enough to explain the stronger correlation observed in the Tiles scheme for the $z_{ph}$ sample and the All scheme for both samples. Therefore, we studied additional sources of surface density variations in the foreground samples.

\subsubsection{Surface density variation of the foreground samples.}
There are different causes of surface density variations in large area galaxy surveys, such as scanning strategy, sensitivity variation with time, and foreground contamination. Moreover, the sample selection can amplify or reduce these variations, for example a region where the conditions for spectroscopic observations are different from the mean field conditions.
The detailed correction of these possible variations is complicated and requires a deep knowledge of the particular details of the instrument and the pipeline used for the production of the catalogue.

For the purpose of this work we adopted a simple approach to investigate the existence and correction of surface density variations in the foreground samples. As we can only observe a discrepancy at the largest angular scales, we decided to focus just on this range.

First, we created a surface density map by adding $+1$ to the pixel value at the position of each galaxy on the sample. Then we smoothed the map using a Gaussian kernel with a certain standard deviation (see discussion later in this section). Next, we applied the H-ATLAS survey masks so that we could neglect border effects due to the smoothing step. These surface density maps are then used to generate the Random catalogues, $R_2$, for the foreground samples (used to estimate $D_1R_2$ and $R_1R_2$ terms in equation \ref{eq:wx}). The bottom panel in Figure \ref{Fig:SD_maps} shows an example of a smoothed surface density map built using the $z_{ph}$ sample, with a standard deviation of 180 arcmin, for the G15 field. As expected, the overall density map at those angular scales is almost homogeneous. However, there are some variations that might be biasing our measurements: the source density in the second Tile from the left is higher than the fourth.

However, the exact value to be used as the Gaussian kernel dispersion is an unknown quantity. Using values smaller than 180 arcmin, the resulting density map starts to mimic the two-halo correlation of the foreground data. This means that the obtained $R_2$ catalogues contain part of the real auto-correlation and remove part of this power from the estimated cross-correlation. For this reason and considering that the cross-correlation function decreases steeply for $\theta \sim 100$ arcmin, we can set a Gaussian dispersion of > 150 arcmin as a lower limit. On the other hand, for dispersion values above 180 arcmin, the surface density variation along the area becomes almost negligible in the derived $R_2$. Therefore, we can considered a dispersion of < 200 -- 220 arcmin as an upper limit. Overall, we decided to proceed using a dispersion of 180 arcmin as a representative value, but taking into account that it is arbitrarily chosen. At the same time, given the uncertainties of the measurements at the relevant angular scales, small variations around the chosen deviation value became only a second order effect in our large-scale measurements.

\vspace{5mm}
When both surface density variations are taken into account to generate the random catalogues the large-scale bias observed in the Tiles scheme for the $z_{ph}$ sample or the All field area one for both samples disappear. 

The right panel of Figure \ref{Fig:xcorr_data} shows the estimated cross-correlation functions using different tiling area schemes for the two samples after all the large-scale bias corrections. The difference between the mean values at each angular scale is much smaller than the uncertainties. Considering this good agreement, we are confident that the measurements can be considered robust in all the angular scales commonly used for the cosmological analysis.

As a final summary, to minimise the number of corrections applied to the data, we recommend applying just the IC correction to the mini-Tile measurements for both samples and to the Tile measurement in the $z_{spec}$ case. In the other cases, it is most relevant to consider the surface density correction.

\section{Cosmological constraints}
\label{sec:results}
Once the cross-correlation measurements are corrected for the different large-scale biases discussed in the previous section, we focus our analysis in their application to the estimate of some relevant parameters as done in BON20: the astrophysical parameters ($M_{min}$, $M_1$ and $\alpha$) and the cosmological parameters ($\Omega_m$, $\sigma_8$ and $h$).

The higher number of pairs in the All tiling scheme should provide the smaller uncertainties at the largest angular scales. However, we are left with only four different regions to minimise the cosmic variance, that, as mentioned before, is the most important source of uncertainty at those angular scales. Moreover, the large-scale bias corrections that have to be applied in this case are not completely objective and can introduce a final bias to the cosmological parameters.
For these reasons and considering the almost perfect agreement between the All tiling scheme and the Tiles schemes for both samples, we decided to maintain just the second case in order to simplify the discussion. 
Therefore, we focus on just four cases, all of which are corrected for the relevant large-scale biases: mini-Tiles and Tiles tiling schemes for both samples ($z_{spec}$ and $z_{ph}$). Both tiling schemes have a much higher number of independent smaller sky areas to try to minimise the error contribution given by the cosmic variance resulting in smaller uncertainties (see right panel of Figure \ref{Fig:xcorr_data}).

\begin{table*}[h] 
\caption{Results obtained from the $z_{spec}$ cross-correlation data sets (the mini-Tiles and Tiles). From left to right, the columns are the parameters, the priors and the results (the mean, $\mu$ with the upper and lower limit at the 68 \% CL, the $\sigma$ and the peak of the posterior distribution) for each data set. Those parameters without a value indicates they are unconstrained, that is that there is no constraint at 68\% CL.}. 
\label{Tab:zspec} 
\centering 
\begin{tabular}{c c c c c c c c} 
\hline 
\hline 
Params & Priors & \multicolumn{3}{c}{mini-Tiles} & \multicolumn{3}{c}{Tiles} \\ 
 & $\mathcal{U}$[a,b] & $\mu$ & $\sigma$ & peak & $\mu$ & $\sigma$ & peak\\ 
 &  & $\pm 68 CL$ & $  $ &  & $\pm 68 CL$ &  & \\ 
\hline 
\\   
$\log(M_{min}/M_\odot)$ & [12.0, 14.0]  & $12.57_{- 0.17}^{+ 0.23}$ &  0.20& 12.61 & $12.61_{- 0.15}^{+ 0.19}$ &  0.18 & 12.56\\ 
\\   
$\log(M_1/M_\odot)$ & [12.5, 15.5] & $14.26_{- 0.38}^{+ 1.24}$ &  0.78& 15.03 & $14.37_{- 0.37}^{+ 1.13}$ &  0.74 & 14.71\\
\\ 
$\alpha$ & [0.5, 1.5] & -- &  -- &  -- &  -- &  -- &  --\\  
\\ 
$\Omega_m$ & [0.1, 0.8] & $ 0.45_{- 0.21}^{+ 0.13}$ &  0.16&  0.38 & $ 0.42_{- 0.24}^{+ 0.14}$ &  0.18 &  0.31\\  
\\
$\sigma_8$ & [0.6, 1.2] & $ 0.84_{- 0.18}^{+ 0.11}$ &  0.14&  0.83 & $ 0.82_{- 0.20}^{+ 0.08}$ &  0.14 &  0.75\\
\\
$h$ & [0.5, 1.0] & -- &  -- &  -- & -- &  -- &  -- \\ 
\\   
\hline 
\hline 
\end{tabular} 
\end{table*} 

\begin{table*}[h] 
\caption{Results obtained from the $z_{ph}$ cross-correlation data sets (the mini-Tiles and the Tiles). From left to right, the columns are the parameters, the priors, and the results (the mean, $\mu$ with the upper and lower limit at the 68 \% CL, the $\sigma$ and the peak of the posterior distribution) for each data set. Those parameters without a value indicates they are unconstrained, that is that there is no constraint at 68\% CL.}
\label{Tab:zph} 
\centering 
\begin{tabular}{c c c c c c c c} 
\hline 
\hline 
Params & Priors & \multicolumn{3}{c}{mini-Tiles} & \multicolumn{3}{c}{Tiles} \\ 
 & $\mathcal{U}$[a,b] & $\mu$ & $\sigma$ & peak & $\mu$ & $\sigma$ & peak\\ 
 &  & $\pm 68 CL$ & $  $ &  & $\pm 68 CL$ &  & \\ 
\hline 
\\   
$\log(M_{min}/M_\odot)$ & [12.0, 14.0]  & $12.60_{- 0.13}^{+ 0.20}$ &  0.18& 12.67 & $12.61_{- 0.13}^{+ 0.20}$ &  0.17 & 12.66\\ 
\\   
$\log(M_1/M_\odot)$ & [12.5, 15.5] & $13.81_{- 1.09}^{+ 0.53}$ &  0.76& 13.60 & $13.95_{- 0.95}^{+ 0.74}$ &  0.76 & 13.74\\
\\ 
$\alpha$ & [0.5, 1.5] & $0.96_{-0.46}^{+0.15}$ &  0.27 &  0.77 &  $0.96_{-0.46}^{+0.15}$ &  0.28 &  0.73\\  
\\ 
$\Omega_m$ & [0.1, 0.8] & $ 0.46_{- 0.18}^{+ 0.11}$ &  0.14&  0.38 & $ 0.46_{- 0.19}^{+ 0.12}$ &  0.15 &  0.39\\  
\\
$\sigma_8$ & [0.6, 1.2] & $ 0.99_{- 0.11}^{+ 0.12}$ &  0.11&  0.98 & $ 0.98_{- 0.10}^{+ 0.16}$ &  0.12 &  1.00\\
\\
$h$ & [0.5, 1.0] & $ 0.71_{- 0.21}^{+ 0.06}$ &  0.14 &  0.50 & -- &  -- &  -- \\ 
\\   
\hline 
\hline 
\end{tabular} 
\end{table*} 

\begin{figure}[ht]
\centering
\includegraphics[width=\columnwidth]{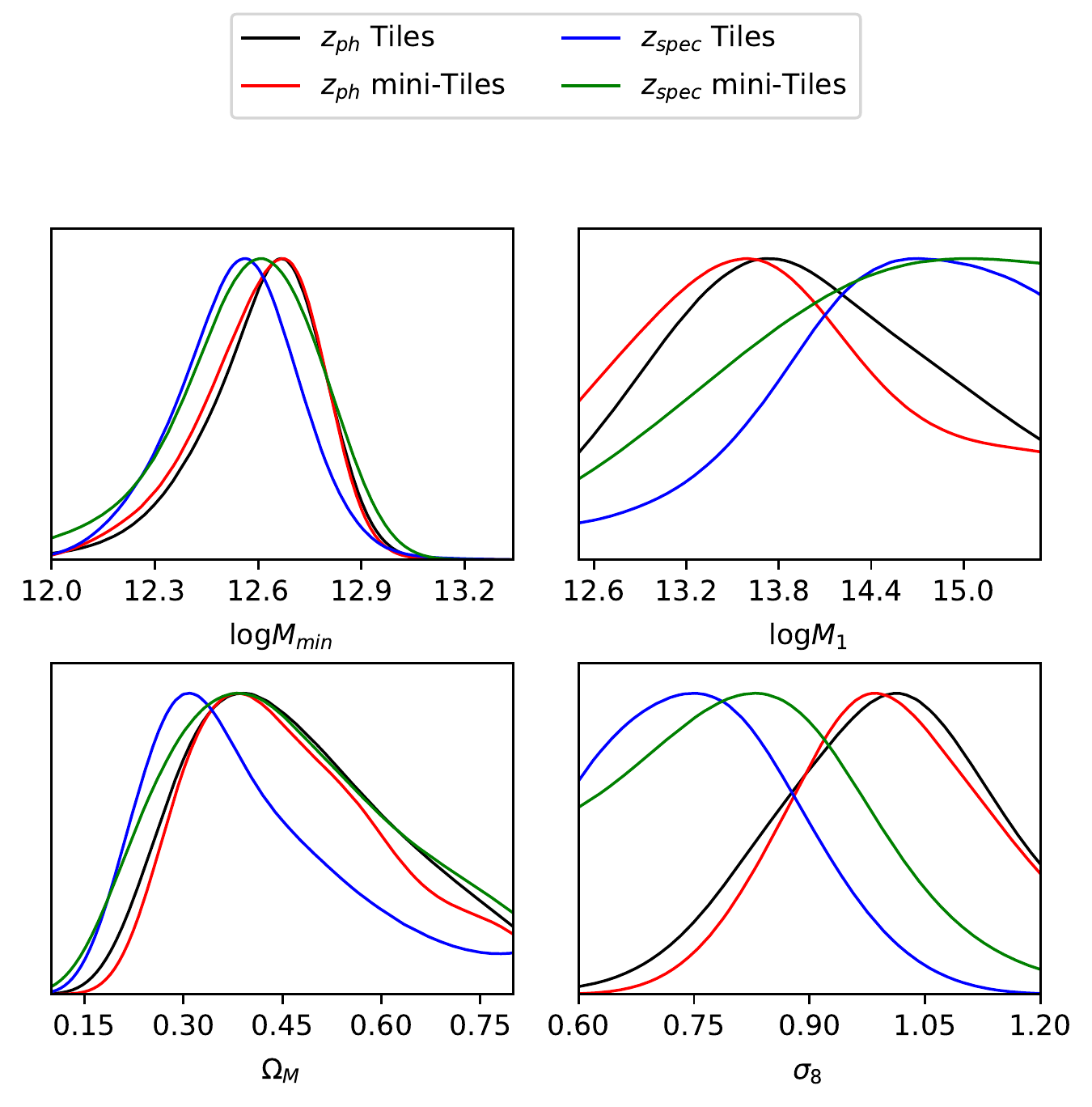}
    \caption{Comparison of the derived posterior distributions for the constrained parameters using the four data sets: $\log M_{min}$ (top left panel), $\log M_1$ (top right), $\Omega_m$ (bottom left), and $\sigma_8$ (bottom right).
 }
    \label{Fig:Comp_1D}
\end{figure}

\subsection{Flat priors}
\label{sec:flat_priors}
The main results of this work are derived by imposing flat priors as described in section \ref{sec:par_estimate}. The full set of posterior distributions can be found in the Appendix \ref{sec:corner_plots}. Figures \ref{Fig:zspec_corner} and \ref{Fig:zph_corner} compare the results derived from the different tiling schemes for the same sample. Moreover, the main statistical quantities that describe the posterior distributions are summarised in Table \ref{Tab:zspec} and \ref{Tab:zph} for $z_{spec}$ and $z_{ph}$ samples, respectively. The model prediction using the best-fit values for both samples using the mini-Tile scheme are shown in the right panel of Figure \ref{Fig:xcorr_data}.

As in BON20 both $\alpha$ and $h$ are not well constrained. For this reason the comparison focuses on the rest of the parameters (see Figure \ref{Fig:Comp_1D}). All the cases provide similar constraints for $M_{min}$ and $\Omega_m$. In the case of $M_{min}$, all of the cases agree to a mean value of $\log(M_{min}/M_\odot) \simeq 12.6 \pm 0.2$ at 68\% CL. This value is very similar with that found by BON20, $\log(M_{min}/M_\odot)= 12.53^{+0.29}_{-0.16}$. With respect the BON20 results, the introduction of the IC correction did not affect the estimated value of this well-constrained parameter. 

In the case of $\Omega_m$, the new results moved the mean, $\sim0.45$, towards lower, more traditional values. This indicates that the large-scale corrections helped to increase slightly the recovered values at the largest angular scales and to reduce their uncertainties. As a consequence, the highest $\Omega_m$ values become less probable based on our current measurements. However, lower limits similar to those found in BON20 are confirmed (e.g. >0.22 for the $z_{spec}$ cases).

On the other hand, the results for $\log M_1$ and $\sigma_8$ are different depending on the sample used. However, the results based on the same sample that use different tile schemes are consistent between them. 

For $M_1$, using the $z_{spec}$ sample, we find a preference for $\log(M_1/M_\odot) \geq 13.8,$ but only at 68\% CL, whereas it shows a clear peak around $\log(M_1/M_\odot) \sim 13.6-13.7$, using the $z_{ph}$ sample. In both cases these results are consistent with those of BON20. In a similar way, $\sigma_8$ mean estimated value moves from $\sim 0.8$, obtained with the $z_{spec}$ sample, to $\sim 1.0$ using the $z_{ph}$ sample. Therefore, with the $z_{spec}$ sample, as in BON20, we obtain similar $\sigma_8$ constraints, but these are not confirmed by the $z_{ph}$ contraints.

Taking into account that the measurements of the cross-correlation function are almost the same between both samples (see again right panel of Figure \ref{Fig:xcorr_data}), this discrepancy in some of the recovered parameters can only be related to the fact that both samples have different redshift distributions. \citet{GON17} performed a tomographic analysis of the cross-correlation function using four different redshift bins, between $0.1<z<0.8$, and study the evolution of the same HOD parameters. While the $M_1$ parameter remains almost constant with redhsift, there is a clear evolution of an increasing $M_{min}$ values with redshift. The results of $\alpha$ are inconclusive as this quantity is unconstrained in most of the redshift bins. By using a single wide redshift bin, we derive an average of the astrophysical parameters weighted by the sample redshift distribution. Therefore, by analysing samples with different redshift distributions, we expect to estimate different astrophysical parameter values, at least for those showing an evolution with redshift as $M_{min}$.

\subsection{Gaussian priors for the unconstrained parameters}
As discussed in the previous section, the two parameters that remain unconstrained with the current data sets are $\alpha$ and $h$. In this section, we study the potential improvements on the results by assuming external constraints on these two parameters. This additional information is introduced in the MCMC as Gaussian priors. For all the analysis in this section we only used the $z_{spec}$ sample with the mini-Tile scheme.

In the case of $\alpha$ we adopted a normal distribution with mean 1.0 and a dispersion of 0.1 (very similar to the Gaussian priors also used in BON20). The results are summarised in Table \ref{Tab:Galpha} and the derived posterior distribution are shown in Figure \ref{Fig:zspec_corner_Galpha}. In general, adopting a Gaussian prior for the $\alpha$ parameters produces almost no variation with respect to the default case. Only the most related parameters, $\log M_1$ and $\sigma_8$, move slightly towards lower values with a reduction in their dispersion of $\sim9$ and $\sim 21$ \%, respectively.

For the Hubble constant, we adopted the two popular values given by the local estimation, $74.03\pm 1.42$ km/s/Mpc \citep{RIE19}, and the CMB value, $67.4\pm0.5$ km/s/Mpc \citep{PLA18_VIII}. The results obtained in these two cases are summarised in Table \ref{Tab:h_high}, while the derived posterior distributions are compared in Figure \ref{Fig:zspec_corner_htest}. The only relevant variation with respect to the default case is that the $\sigma_8$ distribution again moves slightly towards lower values with a reduction on their dispersion of $\sim 29$ \%.

When comparing between both $h$ priors cases, the results are almost identical. However, as also indicated in BON20, higher values of $h$ seem to perform slightly better: the $\Omega_m$ posterior distribution becomes thinner and moves towards lower, more traditional, values. However, the current uncertainties do not allow us to derive stronger conclusions on this particular topic.

Overall, adopting more restrictive priors on the unconstrained parameters does not remarkably improve the results in general. The parameter  $\sigma_8$ seems to benefit more from the reduction of uncertainty in both cases. This is probably because this parameter mostly depends on the intermediate angular scales and, therefore, it is the one mostly affected by changes induced both by the smallest scales (the main influence of $\alpha$ ) and by the largest scales (the main influence of $h$); see appendix in BON20.

\begin{table}[h] 
\caption{Results obtained from the $z_{spec}$ cross-correlation data set using the mini-Tiles scheme, but assuming a Gaussian prior for the $\alpha$ parameter. From left to right, the columns are the parameters, the priors ($\mathcal{U}$[a,b] for Uniform priors, and $\mathcal{N}$[$\mu$,$\sigma$] for the Normal priors) and the results (the mean, $\mu$ with the upper and lower limit at the 68 \% CL, the $\sigma$, and the peak of the posterior distribution).}. 
\label{Tab:Galpha} 
\centering 
\begin{tabular}{c c c c c} 
\hline 
\hline 
Params & Priors & $\mu$ & $\sigma$ & peak\\ 
 & & $\pm 68 CL$ & $  $ &  \\ 
\hline 
\\   
$\log(M_{min}/M_\odot)$ & $\mathcal{U}$[12.0, 14.0]  & $12.53_{- 0.04}^{+ 0.16}$ &  0.21& 12.59 \\ 
\\   
$\log(M_1/M_\odot)$ & $\mathcal{U}$[12.5, 15.5] & $14.31_{- 0.38}^{+ 0.47}$ &  0.71& 14.32\\
\\ 
$\alpha$ & $\mathcal{N}$[1.0, 0.1] & $0.99_{-0.05}^{+ 0.06}$ &  0.10 &  1.00 \\  
\\ 
$\Omega_m$ & $\mathcal{U}$[0.1, 0.8] & $ 0.46_{- 0.16}^{+ 0.01}$ &  0.15&  0.37\\  
\\
$\sigma_8$ & $\mathcal{U}$[0.6, 1.2] & $ 0.76_{- 0.16}^{+ 0.01}$ &  0.10&  0.64\\
\\
$h$ & $\mathcal{U}$[0.5, 1.0] & $ 0.75_{- 0.09}^{+ 0.09}$ &  0.14 &  0.66 \\ 
\\   
\hline 
\hline 
\end{tabular} 
\end{table} 

\begin{table*}[h] 
\caption{Results obtained from the $z_{spec}$ cross-correlation data set using the mini-Tiles IC scheme but assuming two different values for the Hubble constant. The first column is the name of the parameters and then, for each case, the columns are, from left to right, the priors ($\mathcal{U}$[a,b] for Uniform priors, and $\mathcal{N}$[$\mu$,$\sigma$] for the Normal priors), and the results (the mean, $\mu$ with the upper and lower limit at the 68 \% CL, the $\sigma,$ and the peak of the posterior distribution). Those parameters without a value indicate they are unconstrained, that is that there is no constraint at 68\% CL.} 
\label{Tab:h_high} 
\centering 
\begin{tabular}{c c c c c c c c c} 
\hline 
\hline 
Params & \multicolumn{4}{c}{$H_0=74\,km/s/Mpc$} & \multicolumn{4}{c}{$H_0=67\,km/s/Mpc$} \\ 
 & Priors & $\mu$ & $\sigma$ & peak & Priors & $\mu$ & $\sigma$ & peak\\ 
 & & $\pm 68 CL$ & $  $ & & & $\pm 68 CL$ & &  \\ 
\hline 
\\   
$\log(M_{min}/M_\odot)$ & $\mathcal{U}$[12.0, 14.0]  & $12.54_{- 0.05}^{+ 0.13}$ &  0.19& 12.58 & $\mathcal{U}$[12.0, 14.0]  & $12.57_{- 0.06}^{+ 0.12}$ &  0.19& 12.61 \\ 
\\   
$\log(M_1/M_\odot)$ & $\mathcal{U}$[12.5, 15.5] & $14.29_{- 0.01}^{+ 1.02}$ &  0.76& 14.83 & $\mathcal{U}$[12.5, 15.5] & $14.29_{- 0.01}^{+ 1.12}$ &  0.77& 14.93\\
\\ 
$\alpha$ & $\mathcal{U}$[0.5, 1.5] & -- &  -- &  -- & $\mathcal{U}$[0.5, 1.5] & -- &  -- &  -- \\  
\\ 
$\Omega_m$ & $\mathcal{U}$[0.1, 0.8] & $ 0.44_{- 0.15}^{+ 0.01}$ &  0.15&  0.35 & $\mathcal{U}$[0.1, 0.8] & $ 0.49_{- 0.15}^{+ 0.02}$ &  0.15&  0.41\\  
\\
$\sigma_8$ & $\mathcal{U}$[0.6, 1.2] & $ 0.75_{- 0.11}^{+ 0.01}$ &  0.10&  0.69 & $\mathcal{U}$[0.6, 1.2] & $ 0.76_{- 0.15}^{+ 0.01}$ &  0.10&  0.68\\
\\
$h$ & $\mathcal{N}$[0.74, 0.014] & $ 0.74_{- 0.01}^{+ 0.01}$ &  0.014 &  0.74 & $\mathcal{N}$[0.67, 0.005] & $ 0.67_{- 0.003}^{+ 0.002}$ &  0.005 &  0.67\\ 
\\   
\hline 
\hline 
\end{tabular} 
\end{table*} 

\subsection{Combining both data sets}
\label{sec:tomo}

\begin{figure*}[ht]
\centering
\includegraphics[width=\textwidth]{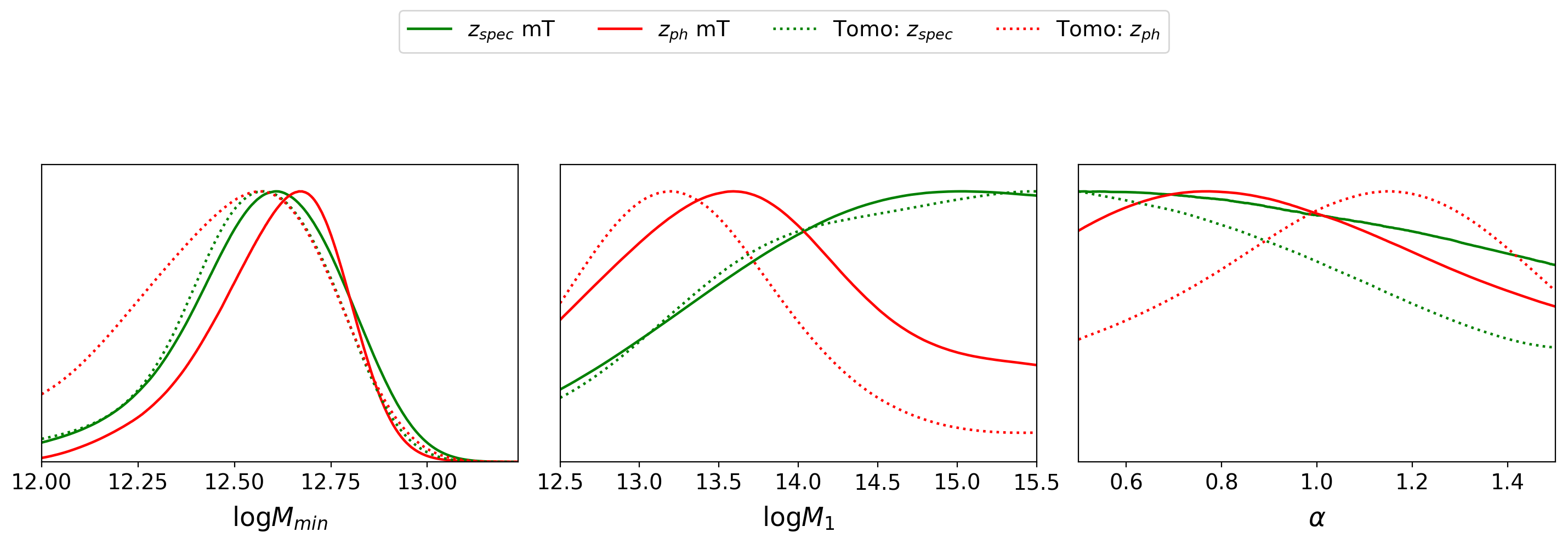}
    \caption{Comparison of the posterior distributions for the astrophysical parameters, $\log( M_{min}/M_\odot)$, $\log( M_{1}/M_\odot)$ and $\alpha$, derived from both samples, $z_{spec}$ and $z_{ph}$, using the mini-Tiles scheme (solid lines) and when combined in a tomographic analysis (dotted lines).
 }
    \label{Fig:Comb_astro}
\end{figure*}

\begin{table}[h] 
\caption{Results obtained from combining both sample, the $z_{spec}$ and the $z_{ph}$, cross-correlation data sets using the mini-Tiles scheme. The adopted priors are the same as in Tables \ref{Tab:zspec} or \ref{Tab:zph}. From left to right, the columns are the parameters and the results (the mean, $\mu$ with the upper and lower limit at the 68 \% CL, the $\sigma,$ and the peak of the posterior distribution).} 
\label{Tab:tomo} 
\centering 
\begin{tabular}{c c c c} 
\hline 
\hline 
Params  & $\mu$ & $\sigma$ & peak\\ 
 & $\pm 68 CL$ & $ $ &  \\ 
\hline 
\\   
$\log(M_{min}/M_\odot)$ $z_{spec}$ & $12.55_{- 0.17}^{+ 0.21}$ &  0.19& 12.58 \\ 
\\   
$\log(M_1/M_\odot)$ $z_{spec}$ & $14.25_{- 0.42}^{+ 1.25}$ &  0.78& 15.49\\
\\ 
$\alpha$ $z_{spec}$& $0.93_{- 0.43}^{+ 0.12}$ &  0.28 &  0.50 \\  
\\
$\log(M_{min}/M_\odot)$ $z_{ph}$ & $12.50_{- 0.22}^{+ 0.26}$ &  0.22& 12.56 \\ 
\\   
$\log(M_1/M_\odot)$ $z_{ph}$ & $13.50_{- 0.90}^{+ 0.30}$ &  0.65& 13.19\\
\\ 
$\alpha$ $z_{ph}$ & $1.04_{- 0.20}^{+ 0.38}$ &  0.26 &  1.15 \\  
\\
$\Omega_m$ & $ 0.50_{- 0.20}^{+ 0.14}$ &  0.15&  0.39\\  
\\
$\sigma_8$ & $ 0.75_{- 0.10}^{+ 0.07}$ &  0.08&  0.75\\
\\
$h$ & -- &  -- &  -- \\ 
\\   
\hline 
\hline 
\end{tabular} 
\end{table} 

The $z_{ph}$ sample has much better statistics with respect to the $z_{spec}$ sample, but we do not see a relevant improvement in the obtained constraints. In addition, even if the measured cross-correlation function is almost the same, each sample provides different results in some of the studied parameters. This is probably linked to the different redshift functions. In this respect, \citet{GON17} tomographic analysis of the cross-correlation function shows a strong evolution with redshift at least for the $\log M_{min}$ parameter. As explained before, by using a single wide redshift bin, the derived astrophysical parameters are the average of the evolving values measured by \citet{GON17} weighted by the particular sample redshift distribution. As we saw, the different averaged astrophysical parameter values between the two samples also affect the recovered values of some of the cosmological parameters. In particular $\sigma_8$ changes from 0.84 for the $z_{spec}$ sample to 0.99 for the $z_{ph}$ sample.

A proper tomographic analysis is beyond the scope of this paper, but we can try a simple, but interesting, analysis by constraining the cosmological parameters using both samples at the same time. We performed a joint analysis allowing different astrophysical parameters constraints for each sample but keeping the same cosmological parameters.

As both samples share partially the same large-scale volume, the measured cross-correlation functions cannot be considered completely independent. To avoid estimating, and taking into account the potential correlation between them, we restricted the $z_{ph}$ sample to just the NGP zone (not used by the $z_{spec}$ sample), almost halving the area used during the previous part of the work for this sample. We estimated the cross-correlation function and the redshift distribution for the restricted $z_{ph}$ sample to be used in the tomographic analysis again. The mean values of the new cross-correlation function agree with the original cross-correlation function ones. However, the uncertainties, mainly at the largest angular scales, increase by $\sim \sqrt{2}$, owing to the smaller area. For the redshift distribution, we do not notice any relevant differences.
Both samples and their estimated cross-correlation functions are now completely independent and can be combined in a single analysis.

After this process, we ran an additional MCMC analysis but this time there were nine parameters to be constrained (for each sample, three astrophysical and three common cosmological 
parameters). We used the mini-Tile scheme for both samples because\ this    scheme requires the simplest large-scale bias correction. The results are summarised in Table \ref{Tab:tomo} and the derived posterior distributions for the nine parameters are shown in Figure \ref{Fig:zspec_corner_tomo}.

Regarding the astrophysical parameters (see Figure \ref{Fig:Comb_astro}) the main changes of the combined analysis with respect to the individual analysis are the following. Imposing a common cosmological parameters values produces a $\log (M_{min}/M_\odot)$ shift towards slightly lower mean values for both samples (from 12.57 to 12.55 for $z_{spec}$ and from 12.60 to 12.50 for $z_{ph}$). For $z_{spec}$ there is only a reduction in the parameter uncertainty for the $M_{1}$ parameter and the $\alpha$ parameter. However, the recovered values for $z_{ph}$ in the combined analysis show a stronger shift of the mean values in both $M_{1}$ and $\alpha$ parameters. The former moves from $\log (M_{1}/M_\odot)$ 13.81 to 13.50 and the associated uncertainty improves from 0.76 to 0.65, while in the case of $\alpha$ from 0.96 to 1.04 with no relevant improvement in the associated uncertainty. This stronger shift could be in part because the higher uncertainties in the large-scale measurements use only half of the total area with respect the original $z_{ph}$ sample.

With respect the cosmological parameters, $\sigma_8$ is the most improved. Its posterior distribution becomes almost Gaussian with a mean value of $\sigma_8=0.75_{- 0.10}^{+ 0.07}$ and a standard deviation of 0.08. However, for $\Omega_m$ the results remain more or less the same. The main reason for this different behaviour between the two parameters is that $\sigma_8$ is more related to the intermediate angular scales while $\Omega_m$ to the largest angular scales whose uncertainty was most affected by halving the $z_{ph}$ sample available area. Therefore, it is possible that the combined analysis also compensated the higher uncertainty in the $z_{ph}$ measurements, which should have worsened the results for the $\Omega_m$ parameter. Finally, the Hubble constant remains unconstrained.
A more detailed discussion on the $\Omega_m$ and $\sigma_8$ results is presented in the next subsection.

\subsection{Comparison with other results}

\begin{figure*}[ht]
\centering
\includegraphics[width=\columnwidth]{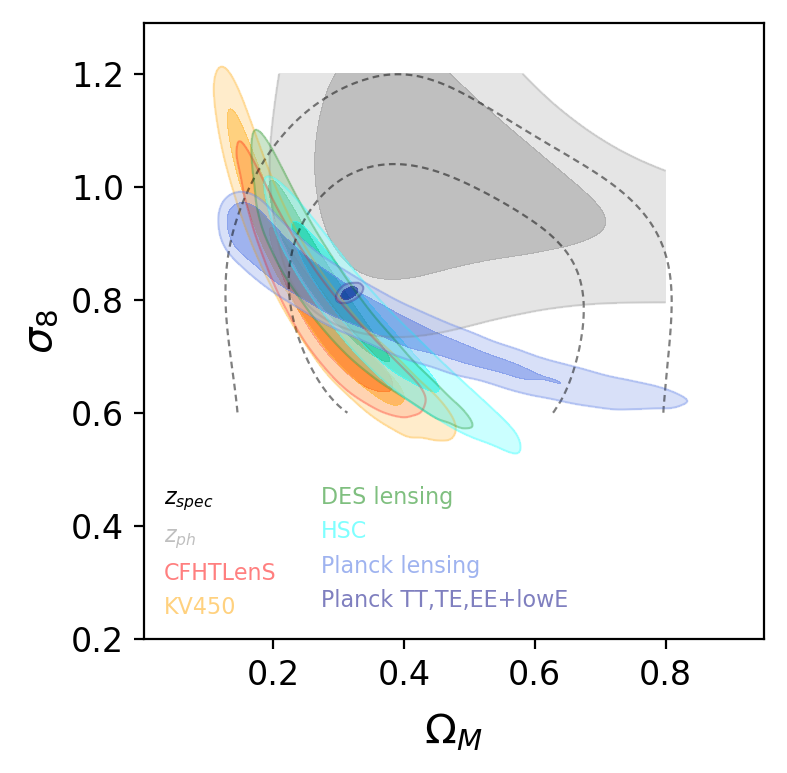}
\includegraphics[width=\columnwidth]{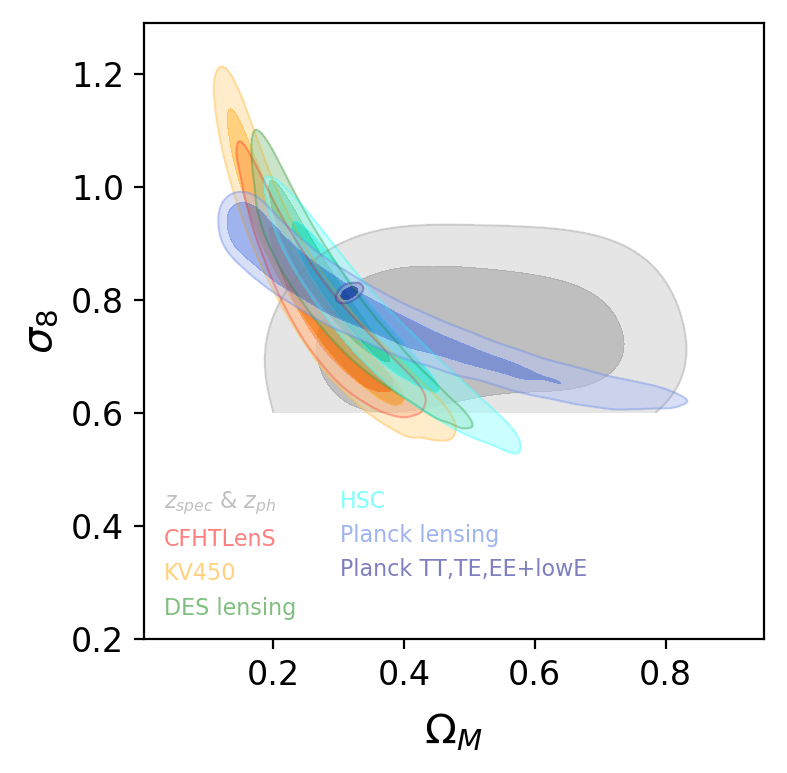}
    \caption{Comparison with external data sets (see text for more details).
 }
    \label{Fig:External_OmgM_sgm8}
\end{figure*}

The weak gravitational lensing results available in the literature are usually related to a different and complementary observable, the shear. In this section we compare with measurements by cosmic shear of galaxies, focussing on the most constraining and the CMB lensing by \textit{Planck} \citep{PLA18_VIII}.
In particular, the results from the following surveys (with different redshift ranges and affected by different systematic effects) are taken into account for the comparison: the Canada-France-Hawaii Telescope Lensing Survey presented in CFHTLenS \citep{JOU17}, the Kilo Degree Survey and VIKING based on $450\,$deg$^2$ data \citep[KV450;][]{HIL20}, the first-year lensing data from the Dark Energy Survey (DES; \cite{TRO18}) and the Subaru Hyper Suprime-Cam first-year data \citep[HSC;][]{HAM20}.

For the comparison, we used the publicly released MCMC results. Moreover, the different results we are comparing with have different priors. Since we are not interested in an in-depth comparison, we do not adjust their priors to our fiducial set-up.

In particular, we compare the constraints in the $\Omega_m$ - $\sigma_8$ plane: cosmic shear measures the combination $\sigma_8 \Omega_m^{0.5}$ and CMB lensing the $\sigma_8 \Omega_m^{0.25}$ combination. Such combinations highlight degeneracy directions, shown in the marginalised posterior contours ($68\%$ and $95\%\,$C.L.) in Figure \ref{Fig:External_OmgM_sgm8} for the data sets described above. For a direct comparison with the literature, the contours of these plots (0.68 and 0.95) are different from those used in the corner plots of this work (0.393 and 0.865, corresponding to the relevant 1-sigma and 2-sigma levels in the 1D histograms in the upper part of the same corner plots).
We also show \textit{Planck} CMB temperature and polarisation angular power spectra, which, although in certain agreement with the HSC and DES constraints, present the tension issues with the CFHTLenS and KV450 data. 

The relevant cosmological constraints derived in this paper are shown in Figure \ref{Fig:External_OmgM_sgm8} for both samples, $z_{spec}$ and $z_{ph}$, using the mini-Tiles scheme. The left panel shows the results from the analysis of each sample individually (grey filled contours for the $z_{ph}$ sample and black dashed curves for the $z_{spec}$ sample), while the right panel shows the results from the combination of both samples as described in section \ref{sec:tomo}. 

With respect to the previous BON20 constraints, by analysing each sample individually, the correction of the large-scale bias has shifted the constraints on the $\Omega_m$ parameter towards lower values, more in agreement with the rest of the results from other studies. 
However, even when combining the two data sets, the Hubble constant remains unconstrained. 

As shown in Figure \ref{Fig:External_OmgM_sgm8}, it is relevant to underline that when both samples are analysed together, the constraints in the $\Omega_m$-$\sigma_8$ plane becomes more restrictive, mainly because $\sigma_8$: $\Omega_m= 0.50_{- 0.20}^{+ 0.14}$ and $\sigma_8= 0.75_{- 0.10}^{+ 0.07}$.
In any case, the constraints derived in this work confirm the main conclusions from BON20. Finally, we note that the data discussed cannot be used to place useful constraints on the Hubble constant yet.

\section{Conclusions}
\label{sec:conclusion}
As discussed in detail in BON20 (see their Figure A.1) the cosmological parameters depend mainly on the largest angular separation measurements. Therefore, the large-scale biases can affect the cosmological constraint derived from the analysis of the magnification bias through the cross-correlation function.

In this work, we study and correct the main large-scale biases that affect our samples to produce a robust estimation of the cross-correlation function. The result is a remarkable agreement among the different cross-correlation measurements, calculated independently of the tiling scheme or foreground samples used. 

Then we analyse these results to estimate cosmological constraints after correcting the different large-scale biases. We get minor improvements with respect to the BON20 results, mainly confirming their conclusions: a lower bound on $\Omega_m > 0.22$ at $95\%$ C.L. and an upper bound $\sigma_8 < 0.97$ at $95\%$ C.L. (results from the $z_{spec}$ sample using the mini-Tile scheme). Therefore, the large-scale biases are a systematic that need to be corrected to derive robust and consistent results between different foreground samples or tiling schemes, but does not help improve 
the precision of the derived constraints much.

In addition, we compare the estimates derived using two different and independent foreground samples: one consisting of foreground galaxies with spectroscopic redshifts, the $z_{spec}$ sample, and another one with only photometric redshifts, the $z_{ph}$ sample. Analysing only one single broad redshift bin, we conclude that the higher errors of the photometric redhsifts do not have a relevant role in our outcomes.
The $z_{ph}$ sample considered in this work has roughly six times more sources than the $z_{spec}$ sample. Its better surface density makes it more sensitive to some large-scale biases, but helps to reduce the uncertainty in the measured cross-correlation function at intermediate and small angular scales. On the other hand, our current results show that the uncertainty is still dominated by the cosmic variance rather than by the surface density of the specific foreground sample at the largest angular scales.

However, the constraints obtained making use of the $z_{ph}$ sample, which provides a more accurate cross-correlation measurements, are generally consistent with those derived using the $z_{spec}$ ones, with similar uncertainties. 

Moreover, adopting Gaussian priors for the unconstrained parameters (i.e. $\alpha$ and the Hubble constant, similarly to BON20) does not improve the results much. Therefore, we are probably reaching the accuracy limit of the cosmological constraints that can be achieved with the analysis of a single redshift bin. Increasing the total area to decrease the cosmic variance even more is probably an interesting improvement to be considered in the future.

Although the measured cross-correlation function is almost the same between both foregrounds samples, we find different constraints for $\log M_1$ and $\sigma_8$ parameters. This is caused by the different redshift distributions between both samples. With a single wide redshift bin, the derived astrophysical parameters, which evolve with time as shown in the tomographic analysis of the cross-correlation function by \citet{GON17}, are averaged quantities weighted by the specific redshift distribution of the selected sample. 

Therefore, we made use of the different redshift distributions to perform a simplified tomographic analysis combining both samples into a single MCMC run. In order to maintain a perfect statistical independence of the measured cross-correlation functions with both samples, we only considered the NGP zone for the $z_{ph}$ sample (not used by the $z_{spec}$ sample). We jointly performed the estimation of the cosmological parameters for both samples, but allowed different values of the astrophysical parameters for each sample. In this way, the effect of having different redshift distributions is included in the astrophysical parameters, allowing us to determine the cosmological parameters with higher precision. The improvements on the $\Omega_m$-$\sigma_8$ plane are evident in the right panel of Figure \ref{Fig:External_OmgM_sgm8}. The cosmological constraints obtained with this independent technique are starting to become competitive with respect to the other lensing results and its particular characteristics make it an interesting possibility in breaking the usual $\Omega_m$-$\sigma_8$ degeneracy.

As a general conclusion, we showed that we are probably reaching the limits of the constraints than can be derived using just a single redshift bin, although there are still some ways to improve the results. However, the most promising advances with the study of the SMGs magnification bias will probably be obtained by performing a more complex tomographic analysis.

\begin{acknowledgements}

JGN, MMC, LB, FA and LT acknowledge the PGC 2018 project PGC2018-101948-B-I00 (MICINN/FEDER). LB and JGN also acknowledge PAPI-19-EMERG-11 (Universidad de Oviedo). 
MM is supported by the program for young researchers ``Rita Levi Montalcini" year 2015. 
A.L. acknowledges support from PRIN MIUR 2017 prot. 20173ML3WW002, `Opening the ALMA window on the cosmic evolution of gas, stars and supermassive black holes', the MIUR grant `Finanziamento annuale individuale attività base di ricerca', and the EU H2020-MSCA-ITN-2019 Project 860744 `BiD4BEST: Big Data
applications for Black hole Evolution STudies'. \\

 We deeply acknowledge the CINECA award under the ISCRA initiative, for the availability of high performance computing resources and support. In particular the projects `SIS19\_lapi', `SIS20\_lapi' in the framework `Convenzione triennale SISSA-CINECA'.\\
 The Herschel-ATLAS is a project with Herschel, which is an ESA space observatory with science instruments provided by European-led Principal Investigator consortia and with important participation from NASA. The H-ATLAS web- site is http://www.h-atlas.org. GAMA is a joint European- Australasian project based around a spectroscopic campaign using the Anglo- Australian Telescope. The GAMA input catalogue is based on data taken from the Sloan Digital Sky Survey and the UKIRT Infrared Deep Sky Survey. Complementary imaging of the GAMA regions is being obtained by a number of independent survey programs including GALEX MIS, VST KIDS, VISTA VIKING, WISE, Herschel-ATLAS, GMRT and ASKAP providing UV to radio coverage. GAMA is funded by the STFC (UK), the ARC (Australia), the AAO, and the participating institutions. The GAMA web- site is: http://www.gama-survey.org/.\\
Funding for the Sloan Digital Sky Survey IV has been provided by the Alfred P. Sloan Foundation, the U.S. Department of Energy Office of Science, and the Participating Institutions. SDSS-IV acknowledges
support and resources from the Center for High-Performance Computing at
the University of Utah. The SDSS web site is www.sdss.org.

SDSS-IV is managed by the Astrophysical Research Consortium for the 
Participating Institutions of the SDSS Collaboration including the 
Brazilian Participation Group, the Carnegie Institution for Science, 
Carnegie Mellon University, the Chilean Participation Group, the French Participation Group, Harvard-Smithsonian Center for Astrophysics, 
Instituto de Astrof\'isica de Canarias, The Johns Hopkins University, Kavli Institute for the Physics and Mathematics of the Universe (IPMU) / 
University of Tokyo, the Korean Participation Group, Lawrence Berkeley National Laboratory, 
Leibniz Institut f\"ur Astrophysik Potsdam (AIP),  
Max-Planck-Institut f\"ur Astronomie (MPIA Heidelberg), 
Max-Planck-Institut f\"ur Astrophysik (MPA Garching), 
Max-Planck-Institut f\"ur Extraterrestrische Physik (MPE), 
National Astronomical Observatories of China, New Mexico State University, 
New York University, University of Notre Dame, 
Observat\'ario Nacional / MCTI, The Ohio State University, 
Pennsylvania State University, Shanghai Astronomical Observatory, 
United Kingdom Participation Group,
Universidad Nacional Aut\'onoma de M\'exico, University of Arizona, 
University of Colorado Boulder, University of Oxford, University of Portsmouth, 
University of Utah, University of Virginia, University of Washington, University of Wisconsin, 
Vanderbilt University, and Yale University. \\
In this work, we made extensive use of \texttt{GetDist} \citep{GETDIST}, a Python package for analysing and plotting MC samples. In addition, this research has made use of the python packages \texttt{ipython} \citep{ipython}, \texttt{matplotlib} \citep{matplotlib} and \texttt{Scipy} \citep{scipy}

\end{acknowledgements}

  \bibliographystyle{aa} 
  \bibliography{./XCORR_ZPH} 

\appendix
\section{Posterior distributions of the MCMC results}
\label{sec:corner_plots}
Posterior distributions for the different analyses discussed during the article. The contours for all these plots are set to 0.393 and 0.865. We note that the relevant 1-sigma and 2-sigma levels for a 2D histogram of samples is 39.3\% and 86.5\% rather than 68\% and 95\%. Otherwise, there is not a direct comparison with the 1D histograms above the contours.

\begin{figure*}[ht]
\centering
\includegraphics[width=\textwidth]{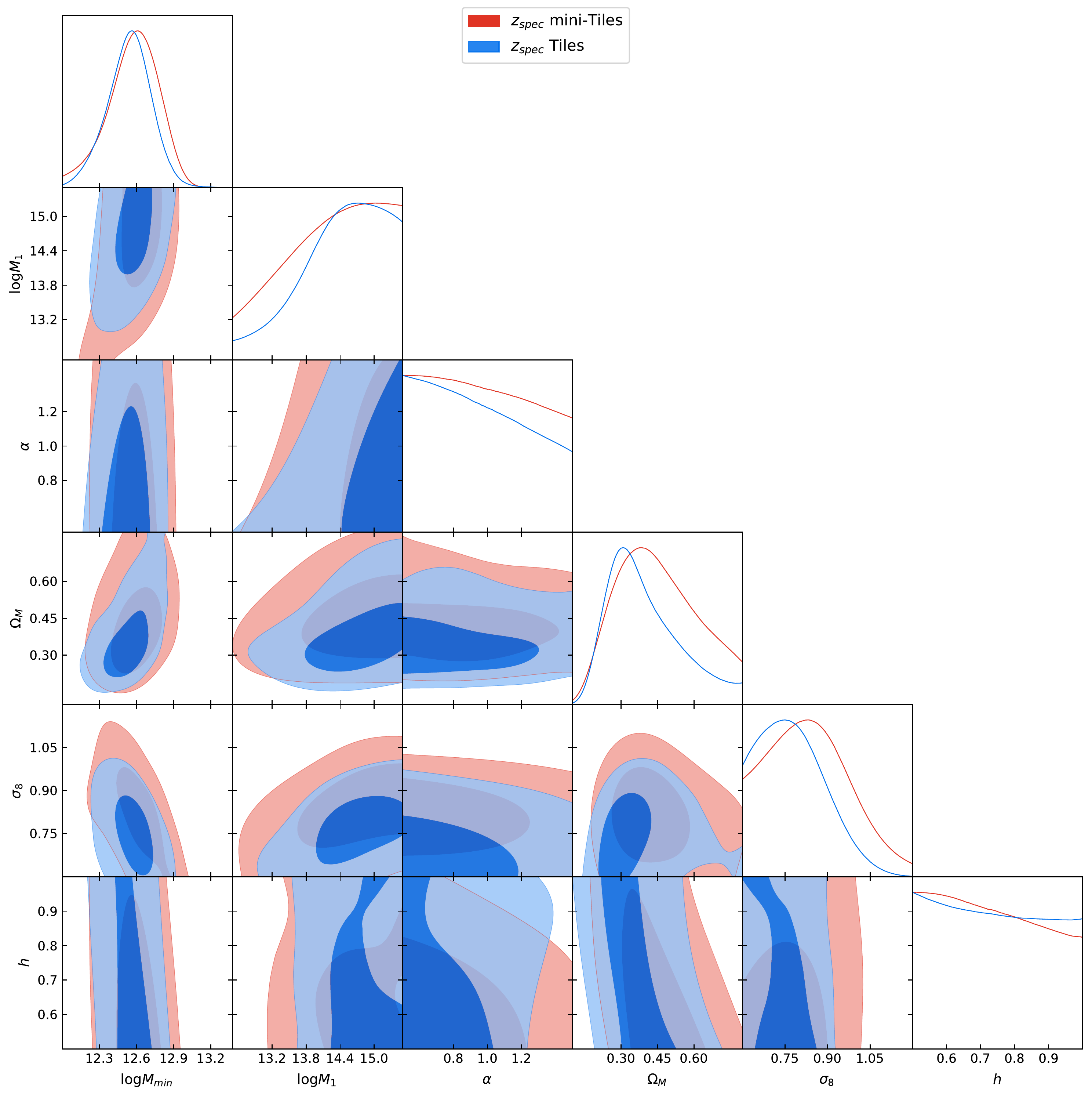}
    \caption{MCMC results for the $z_{spec}$ data sets (the mini-Tiles scheme results in red and the Tiles results in blue).
 }
    \label{Fig:zspec_corner}
\end{figure*}

\begin{figure*}[ht]
\centering
\includegraphics[width=\textwidth]{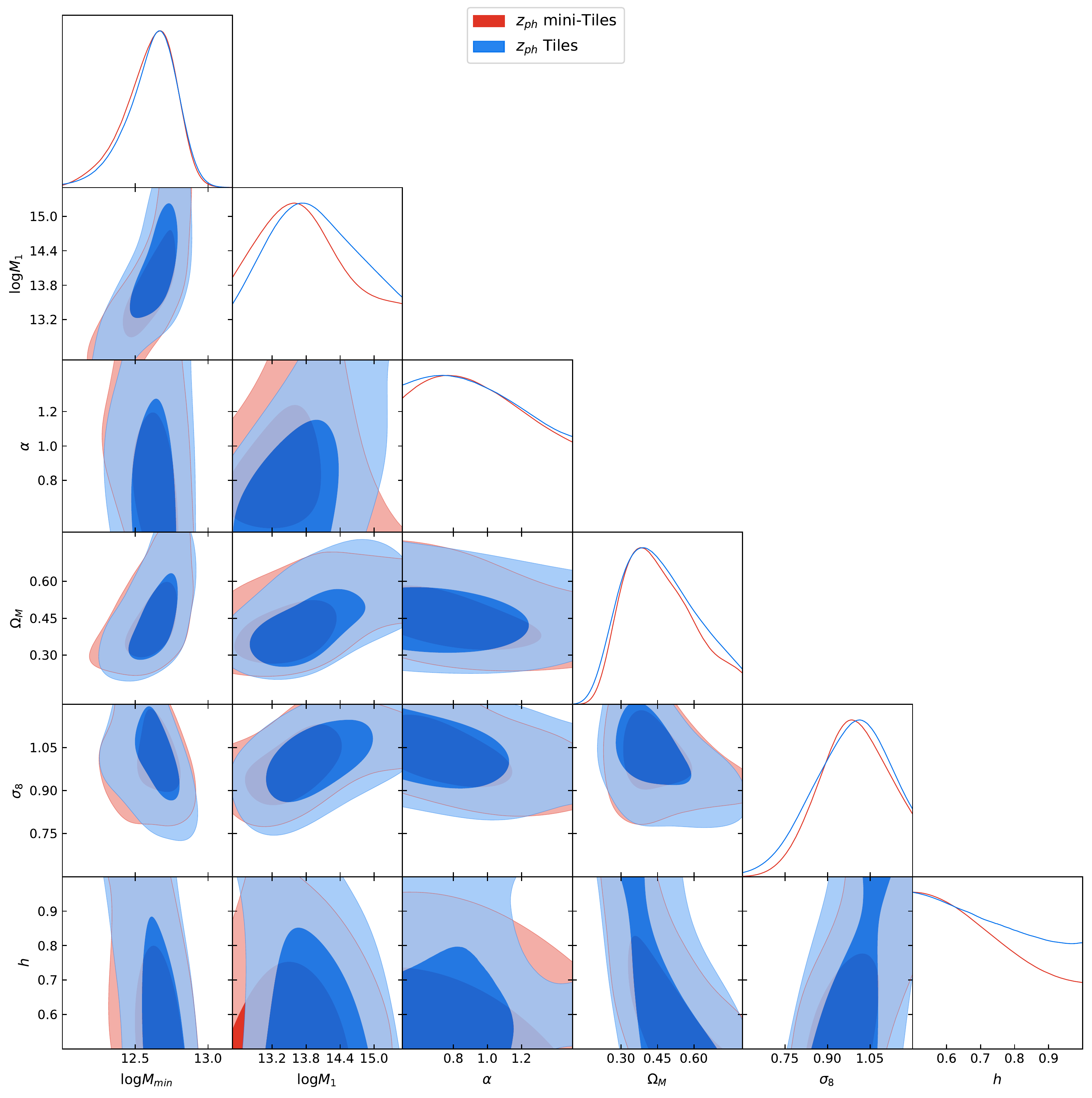}
    \caption{MCMC results for the $z_{ph}$ data sets (the mini-Tiles scheme results in red and the Tiles results in blue).
 }
    \label{Fig:zph_corner}
\end{figure*}

\begin{figure*}[ht]
\centering
\includegraphics[width=\textwidth]{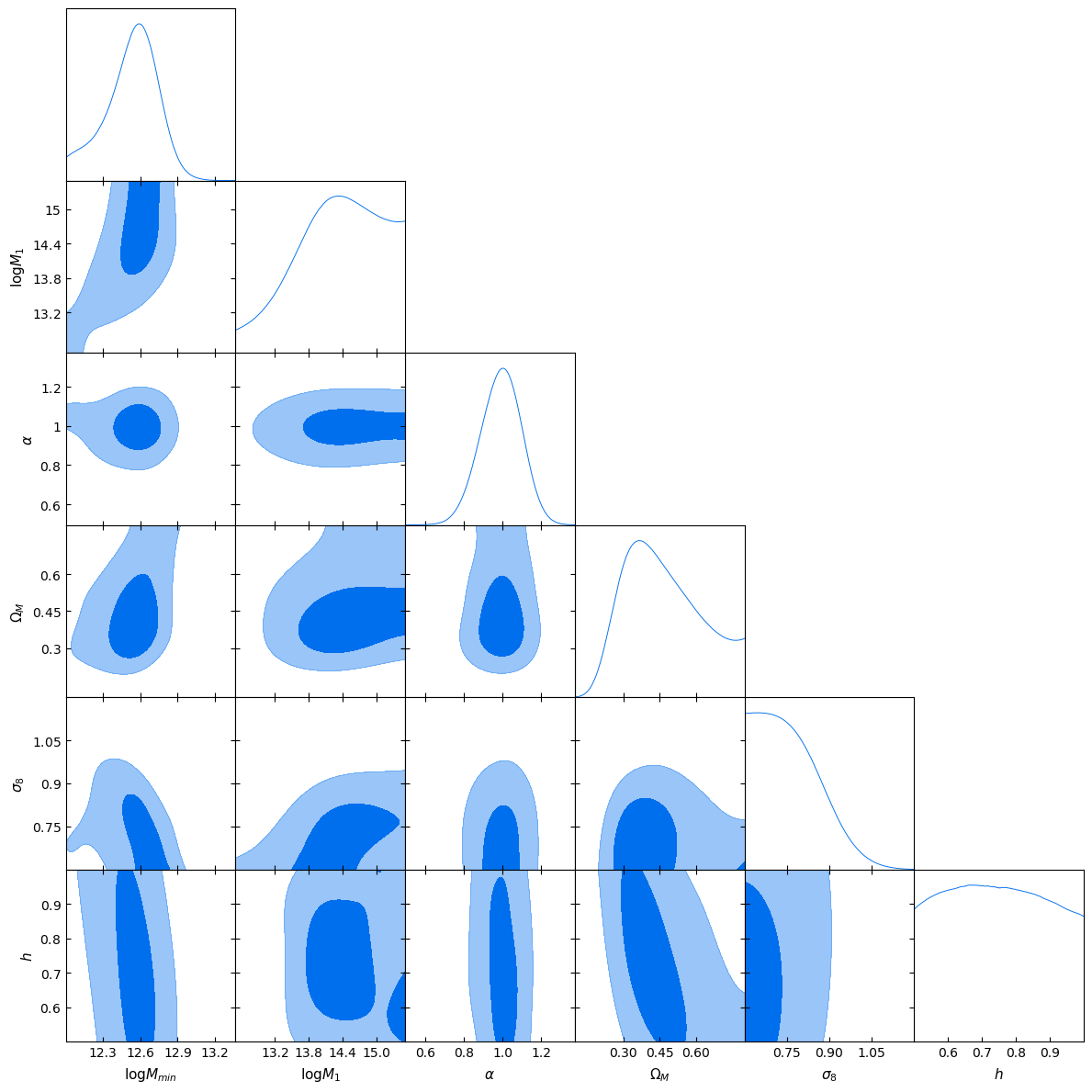}
    \caption{MCMC results for the $z_{spec}$ sample and mini-Tile scheme assuming a Gaussian prior for $\alpha$.
 }
    \label{Fig:zspec_corner_Galpha}
\end{figure*}

\begin{figure*}[ht]
\centering
\includegraphics[width=\textwidth]{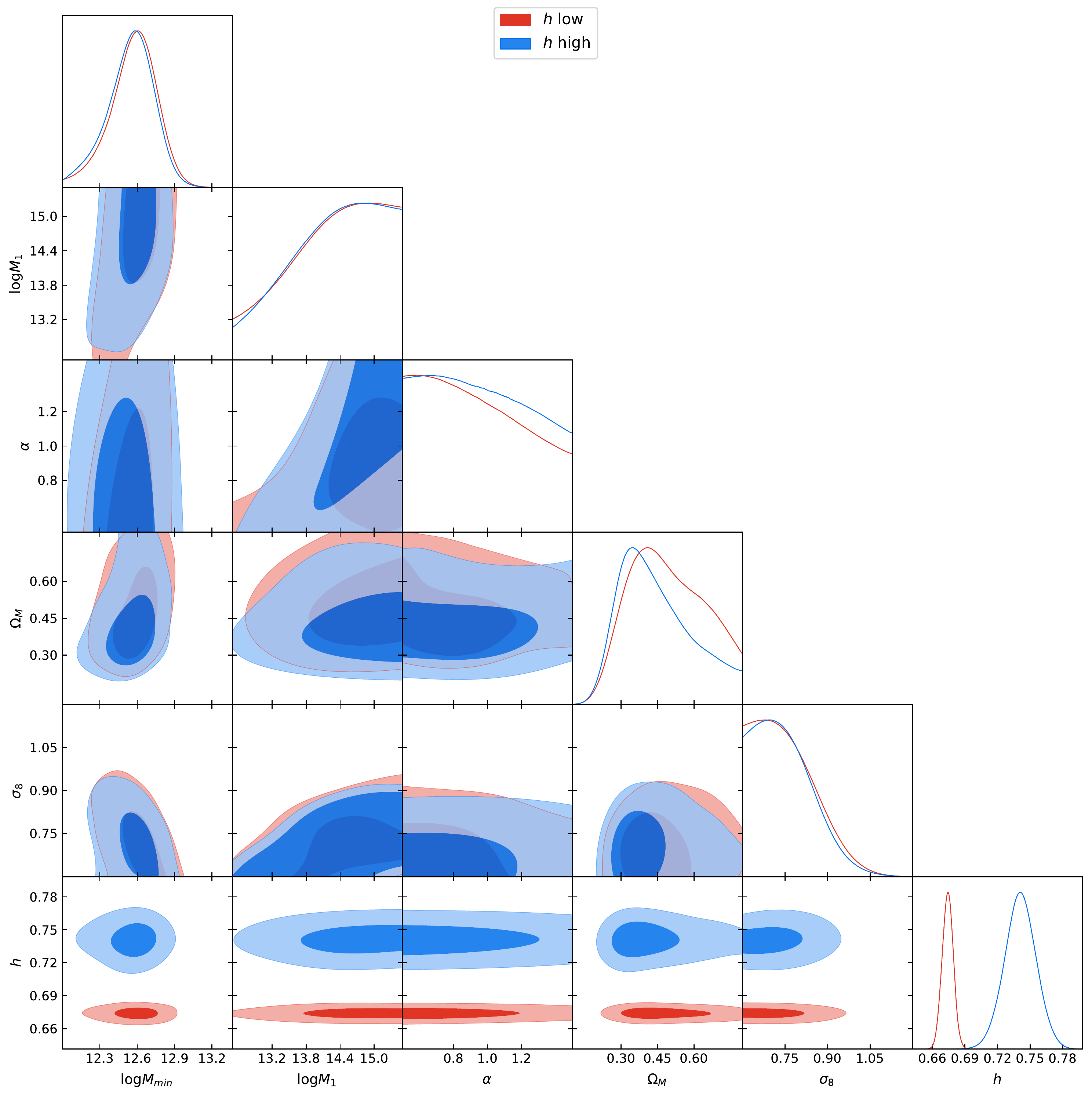}
    \caption{MCMC results for the $z_{spec}$ sample and mini-Tile scheme assuming two different Gaussian priors for $h$. The two popular values given by the local estimation were adopted as follows: $74.03\pm 1.42$ km/s/Mpc \citep[blue;][]{RIE19}, and the CMB value, $67.4\pm0.5$ km/s/Mpc \citep[red;][]{PLA18_VIII}.
 }
    \label{Fig:zspec_corner_htest}
\end{figure*}

\begin{figure*}[ht]
\centering
\includegraphics[width=\textwidth]{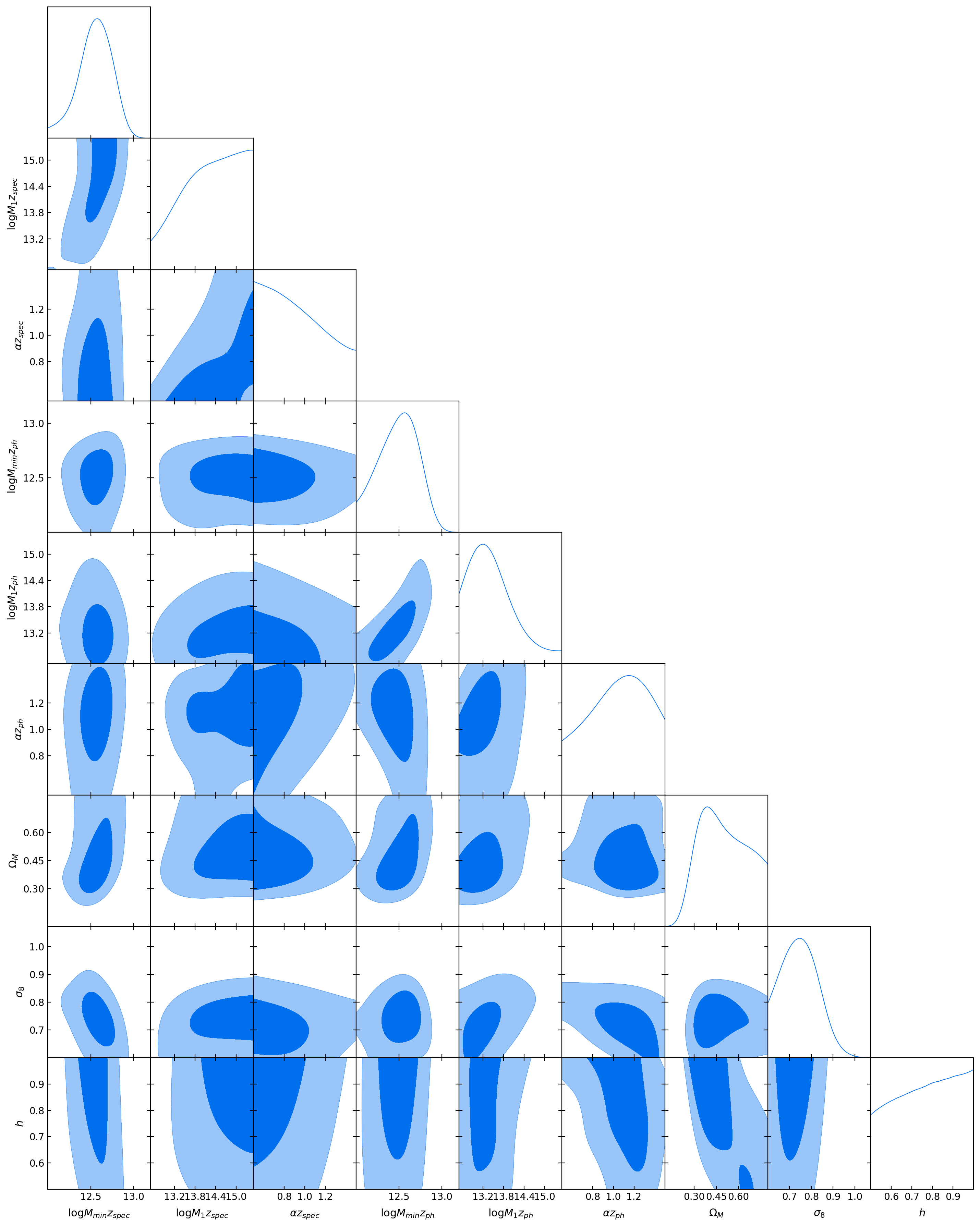}
    \caption{MCMC results combining the $z_{spec}$ and the $z_{ph}$ samples in a tomographic run using the mini-Tile scheme.
 }
    \label{Fig:zspec_corner_tomo}
\end{figure*}

\end{document}